\documentclass[useAMS,usenatbib]{mn2e}

\usepackage{amsmath}
\usepackage{amssymb}
\usepackage{graphicx}
\usepackage{aas_macros}

\title[Importance of initial conditions for SF]{Importance of the Initial Conditions for Star Formation -- III: Statistical Properties of Embedded Protostellar Clusters}
\author[Girichidis et al.]{\Large{Philipp~Girichidis$^{1,2,3}$\thanks{email: \texttt{philipp@girichidis.com}}, Christoph~Federrath$^{4,2}$, Richard~Allison$^2$, Robi~Banerjee$^{1}$, \& Ralf~S.~Klessen$^2$}
\vspace*{0.2cm} \\
\scriptsize
$^1$Hamburger Sternwarte, Gojenbergsweg 112, 21029 Hamburg, Germany\\
\scriptsize
$^2$Zentrum f\"ur Astronomie der Universit\"at Heidelberg, Institut f\"ur Theoretische Astrophysik, Albert-Ueberle-Str.~2, 69120 Heidelberg, Germany \\
\scriptsize
$^3$Cardiff School of Physics and Astronomy, The Parade, Cardiff, CF24 3AA, UK \\
\scriptsize
$^4$Monash Centre for Astrophysics (MoCA), School of Mathematical Sciences, Monash University, Vic 3800, Australia
}

\newcommand{\ColorBar}{
  \vspace{0.2cm}
  \includegraphics[width=16cm]{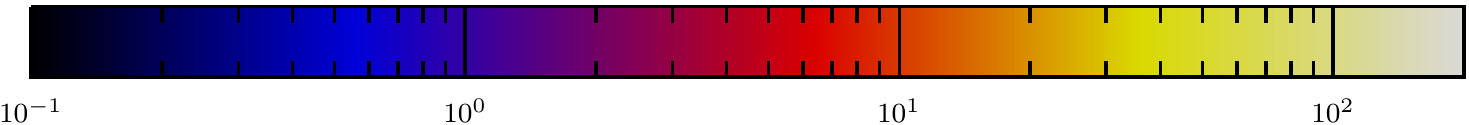}\\
  column density [g~cm$^{-2}$]\\\vspace{0.2cm}
}

\newcommand{\rhoav}{\langle\rho\rangle}
\newcommand{\rkl}[1]{\left(#1\right)}

\newcommand{\skl}[1]{\left\langle#1\right\rangle}

\newcommand{\EkinOverEpotf}{\frac{E_\text{kin}}{|E_\text{pot}|}}
\newcommand{\EthermOverEpotf}{\frac{E_\text{therm}}{|E_\text{pot}|}}
\newcommand{\phn}{\phantom{0}}
\newcommand{\phnn}{\phantom{0}\phantom{0}}

\begin{document}

\maketitle
\begin{abstract}
We investigate the formation of protostellar clusters during the collapse of dense molecular cloud cores with a focus on the evolution of potential and kinetic energy, the degree of substructure, and the early phase of mass segregation. Our study is based on a series of hydrodynamic simulations of dense cores, where we vary the initial density profile and the initial turbulent velocity. In the three-dimensional adaptive mesh refinement simulations, we follow the dynamical formation of filaments and protostars until a star formation efficiency of 20\%. Despite the different initial configurations, the global ensemble of all protostars in a setup shows a similar energy evolution and forms sub-virial clusters with an energy ratio $E_\mathrm{kin}/|E_\mathrm{pot}|\sim0.2$. Concentrating on the innermost central region, the clusters show a roughly virialised energy balance. However, the region of virial balance only covers the innermost $\sim10-30\%$ of all the protostars. In all simulations with multiple protostars, the total kinetic energy of the protostars is higher than the kinetic energy of the gas cloud, although the protostars only contain 20\% of the total mass. The clusters vary significantly in size, mass, and number of protostars, and show different degrees of substructure and mass segregation. Flat density profiles and compressive turbulent modes produce more subclusters then centrally concentrated profiles and solenoidal turbulence. We find that dynamical relaxation and hence dynamical mass segregation is very efficient in all cases from the very beginning of the nascent cluster, i.e., during a phase when protostars are constantly forming and accreting.
\end{abstract}


\section{Introduction}

In the current paradigm of star formation, most of the stars form in a clustered environment \citep{Lada03}. Concerning massive stars, studies by \citet{deWitEtAl2004,deWitEtAl2005} give an upper limit of $4\pm2\%$ of O-stars which could not be traced back to star clusters and which are thus candidates for the isolated formation of massive stars. Further work by \citet{SchilbachRoeser2008}, \citet{GvaramadzeBomans2008}, and \citet{PflammAltenburgKroupa2010} even allow for a lower fraction of O-stars that have to form in a clustered environment. Consequently, an understanding of star formation is ultimately linked to the formation of clusters and stars within them.

Over a huge spatial range of astrophysical objects and thus also during the collapse of a molecular cloud and the formation of a stellar cluster, the observed kinetic energy shows a robust scaling with the size of the object \citep{Larson81,SolomonEtAl1987,Ossenkopf02,Heyer04,HilyBlantEtAl2008,RomanDuvalEtAl2011}. This global analysis, however, does not take into account the spatial and dynamical substructure of small-scale collapsing regions with sizes below $0.1\,\mathrm{pc}$. The energy balance and virial state of the star-forming region may vary during the formation of the cluster and for different degrees of substructure in a cloud. Local changes in the dynamics may lead to different formation modes of the cluster and alter the stellar distribution and the accretion process in a nascent cluster.

Within a cluster, the distribution of stars is generally not uniform, but shows signatures of mass segregation with a tendency of more massive stars to be located closer to the centre of the cluster. This phenomenon is observed in many young clusters \citep{Hillenbrand98,StolteEtAl2005,Stolte06,KimEtAl2006,HarayamaEtAl2008,EspinozaEtAl2009,BontempsEtAl2010,GennaroEtAl2011}. However, the detailed definitions of mass segregation and the regions where they apply lead to controversies. \citet{KirkMyers2011} find mass segregation in small groups in Taurus, \citet{ParkerEtAl2011} find more massive stars to be inversely mass segregated, concentrating on the complex as a whole. In addition, there is a strong debate, whether the observed mass segregation in young clusters is primordial or due to dynamical interactions via two-body relaxation. One fundamental problem of that debate lies within the definition of mass segregation and the methods and tools to determine the segregation state. \citet{Allison09} define mass segregation as massive stars located close to other massive stars. \citet{KirkMyers2011} base their mass segregation on the distance of the more massive stars to the centre of the local association. \citet{MaschbergerClarke2011} investigated mass segregation of a collection of smaller cores and modified the model by \citet{Allison09} to be more robust in case of outlier stars. In addition, they also used local surface density as a measure of mass segregation. Generally, the substructure of the region in question plays a significant role in the explanation of the origin of mass segregation. Whereas the global system might not have enough time to dynamically relax, the small individual subclusters might well be able to reach a relaxed segregated state. In addition, the final mass segregation may crucially depend on how much degree of mass segregation is preserved during the merger of small subclusters, i.e. how much mass segregation the merged structure can inherit from its constituents. Consequently, a combined investigation of the degree of substructure as a function of time, the energetic state of the cloud, the formation mode of stars within the cluster, and the formation of the clusters themselves is absolutely crucial to understand the mass segregation process.

In this study we analyse the dynamical evolution of collapsing cloud cores and their virial state before and during the formation of protostars. In addition, we investigate the resulting substructure during the collapse and the possible degree of dynamical mass segregation for dense collapsing cloud cores in numerical simulations. We vary the initial density profile as well as the initially imposed turbulent motions and analyse their impact on the later cluster structure. The simulations, which are taken from \citet{Girichidis11a}, hereafter Paper~I, follow the collapse of the core and the formation of protostars. We find that the initial conditions have a large impact on the degree of substructure in a cluster and that the clusters show strong dynamical interactions between the protostars. As a result, the individual subclusters are very likely to have enough time for dynamical mass segregation. In contrast, for the global cloud, the time scales for dynamical relaxation are too long in comparison to the time scale at which stars form in these dense cores. Due to the strong dynamical interactions in the central region of the (sub)clusters from their formation onwards, it is basically impossible to define primordial mass segregation in the simulated cores.

The paper is organised as follows. Section~\ref{sec:methods-initial-conditions} covers the description of the simulations and the key properties of the numerical setups. In section~\ref{sec:analysis} we introduce the methods that we use to analyse the energy state, the degree of substructure, and the mass segregation. Section~\ref{sec:results} presents our results, separately for the global cloud and the central or main subclusters. Section~\ref{sec:discussion} and~\ref{sec:conclusion} comprise the discussion of the obtained results and the conclusions, respectively.
 
\section{Numerical Methods \& Initial Conditions}
\label{sec:methods-initial-conditions}
The simulation data used in this study are the same as in Paper~I, where a detailed description of the initial setups can be found. Here we only summarise the key parameters.
 
\subsection{Global Simulation Parameters}
We simulate the collapse of a spherically symmetric molecular cloud with a radius of $R=0.1~\mathrm{pc}$ and a total mass of $100~M_\odot$. The resulting average density is $\rhoav = 1.76\times10^{-18}~\mathrm{g~cm}^{-3}$ and the corresponding free-fall time gives $t_\mathrm{ff}=5.02\times10^4~\mathrm{yr}$. The gas with a mean molecular weight of $\mu=2.3$ is assumed to be isothermal at a temperature of $20~\mathrm{K}$, yielding a constant sound speed of $c_\mathrm{s}=2.68\times10^4~\mathrm{cm~s}^{-1}$. The Jeans length, $\lambda_\mathrm{J}$, and the corresponding Jeans mass $M_\mathrm{J}$, calculated as a sphere with diameter $\lambda_\mathrm{J}$, are $\lambda_\mathrm{J}=9300~\mathrm{AU}$ and $M_\mathrm{J}=1.23~M_\odot$, respectively. Table~\ref{tab:phys-param} provides an overview of all physical parameters.

\begin{table}
  \caption{Physical parameters of all setups}
  \label{tab:phys-param}
  \begin{tabular}{lcc}
    Parameter & & Value\\
    \hline
    cloud radius & $R_0$ & $3\times10^{17}\,\mathrm{cm}\approx0.097\,\mathrm{pc}$\\
    total cloud mass & $M_\text{tot}$ &$100~M_\odot$\\
    mean mass density & $\rhoav$ & $1.76\times10^{-18}$~g~cm$^{-3}$\\
    mean number density & $\langle n \rangle$ & $4.60\times10^5$~cm$^{-3}$\\
    mean molecular weight & $\mu$ & $2.3$\\
    temperature & $T$ & 20~K\\
    sound speed & $c_\text{s}$ & $0.27$~km~s$^{-1}$\\
    rms Mach number & $\mathcal{M}$ & $3.28-3.64$\\
    mean free-fall time & $t_\text{ff}$ & $5.02\times10^4~\text{yr}$\\
    sound crossing time & $t_\text{sc}$ & $7.10\times10^5~\text{yr}$\\
    turbulent crossing time & $t_\text{tc}$ & $1.95 - 2.16\times10^5~\text{yr}$\\
    Jeans length & $\lambda_\text{J}$ & $9.26\times10^3\,\mathrm{AU}\approx0.23~R_0$\\
    Jeans volume & $V_\text{J}$ & $1.39\times10^{51}~\text{cm}^3$\\
    Jeans mass & $M_\text{J}$ & $1.23~M_\odot$\\
    \hline
  \end{tabular}
\end{table}

\subsection{Numerical Code}
The simulations were carried out with the astrophysical code FLASH Version 2.5 \citep{FLASH00}. To integrate the hydrodynamic equations, we use the piecewise-parabolic method (PPM) by \citet{Colella84}. The computational domain is subdivided into blocks containing a fixed number of cells with an adaptive mesh refinement (AMR) technique based on the PARAMESH library \citep{PARAMESH99}.

\subsection{Resolution and Sink Particles}
The simulations were run with a maximum effective resolution of $4096^3$ grid cells, corresponding to a smallest cell size of $\Delta x\approx13~\mathrm{AU}$. In order to avoid artificial fragmentation, the Jeans length has to be resolved with at least $4$ grid cells \citep{Truelove97}. To resolve turbulence on the Jeans scale, however, a significantly higher number of cells is required. \citet{FederrathEtAl2011} find a minimum resolution of about 30 cells per Jeans length. Due to the high computational demand, we only use 8 cells in the current runs, so we likely miss some turbulent energy in our cores, which provides additional support against gravitational collapse. We might thus slightly overestimate the amount of fragmentation and underestimate the formation times of protostars. It must be noted, however, that this is a general limitation of all present star cluster formation calculations because resolving the Jeans length with more than $10-20$ cells can be computationally prohibitive. Additionally, in order to terminate local runaway collapse in a controlled way, we use sink particles (see e.g., \citealt{Bate95}, \citealt{Krumholz04}, \citealt{Federrath10a}). They are introduced at the highest level of the AMR hierarchy. A necessary but not sufficient criterion for the formation of sink particles is that the gas density needs to be higher than the threshold value
\begin{equation}
  \rho_\text{max} = \frac{\mathrm{\pi} c_\text{s}^2}{4\,G\,(3\,\Delta x)^2} = 2.46\times10^{-14}\text{g~cm}^{-3}.
\end{equation}
If a cell exceeds this density, a spherical control volume with a radius of $3\Delta x$ is investigated for the following gravitational collapse indicators \citep{Federrath10a}: 
  \begin{itemize}
    \item The gas is converging along all principal axis, $x$, $y$, and $z$,
    \item has a central minimum of the gravitational potential,
    \item is Jeans-unstable,
    \item is gravitationally bound, and
    \item is not within the accretion radius of an already existing sink particle.
  \end{itemize}
If the collapse criteria are fulfilled, an accreting Lagrangian sink particle is formed. This sink particle is then identified as an individual protostar \citep{Bate95,WuchterlKlessen2001}. Table~\ref{tab:simul-param} lists the simulation and resolution parameters.

\begin{table}
  \caption{Numerical simulation parameters}
  \label{tab:simul-param}
  \begin{tabular}{lcc}
    Parameter & & Value\\
    \hline
    simulation box size & $L_\text{box}$ & $0.26$~pc\\
    smallest cell size & $\Delta x$ & $13.06$~AU\\
    Jeans length resolution & & $\ge8~(6^*)$ cells\\
    max. gas density & $\rho_\text{max}$ & $2.46\times10^{-14}$~g~cm$^{-3}$\\
    max. number density& $n_\mathrm{max}$ & $6.45\times10^{9}$~cm$^{-3}$\\
    sink particle accretion radius &$r_\text{accr}$ & $39.17$~AU\\
    \hline
  \end{tabular}

  \medskip
  $^*$ at highest level of refinement

\end{table}

\subsection{Initial Conditions}
The following four density profiles were used:
\begin{enumerate}
\item Top-hat profile, $\rho=\mathrm{const}$ (TH)
\item Rescaled Bonnor-Ebert sphere. (BE)
\item Power-law profile, $\rho\propto r^{-1.5}$ (PL15)
\item Power-law profile, $\rho\propto r^{-2.0}$ (PL20).
\end{enumerate}
A detailed  description of the profiles can be found in Paper~I.

The turbulence is modelled with an initial random velocity field, originally created in Fourier space, and transformed back into real space. The power spectrum of the modes is given by a power-law function in wavenumber space ($\mathbf{k}$ space) with $E_\text{k}\propto k^{-2}$, corresponding to Burgers turbulence, consistent with the observed spectrum of interstellar turbulence \citep[e.g.,][]{Larson81,Ossenkopf02,Heyer04}. The velocity field is dominated by large-scale modes due to the steep power-law exponent, $-2$, with the largest mode corresponding to the size of the simulation box. Concerning the nature of the $\mathbf{k}$ modes, compressive (curl-free) modes are distinguished from solenoidal (divergence-free) ones. The simulation uses three types of initial fields: purely compressive fields (c), purely solenoidal (s), and a natural (random) mixture (m) of both. The choice of these different turbulent fields was motivated by the strong impact of the nature of the modes on the cloud evolution, found by \citet{Federrath08, Federrath10b}. Note however that only decaying turbulence with compressive, solenoidal, and mixed modes are considered here.

All setups have supersonic velocities with an rms Mach number
$\mathcal{M} = v_\mathrm{rms}/c_\mathrm{s}$
ranging from $\mathcal{M}=3.28-3.54$ with an average of $\skl{\mathcal{M}}=3.44$.
The sound crossing time through the entire cloud is
$t_\mathrm{sc}(R_0) = 7.10\times10^5~\mathrm{yr}$, and
the time for gas with an average velocity of $\skl{\mathcal{M}}c_\mathrm{s}$ to cross the cloud is
$t_\mathrm{tc}(R_0) = 2.06\times10^5~\mathrm{yr}$,
respectively.

We combine four density profiles with six different turbulent velocity fields (three different compositions of modes with two different random seeds each). Table~\ref{tab:simulation-global-overview} shows a list of all models.

\begin{table*}
  \caption{List of the runs and their main properties}
  \label{tab:simulation-global-overview}
  \begin{tabular}{lcccccccccccc}
    name  & $\mathcal{M}$ & total &  total & $t_\text{sim}$ &$t_\text{sim}/t_\text{ff}$ & $N_\text{sink}$ & $M_\text{max}$ & $n_*^\mathrm{glob}$ & $\skl{s}$& $\skl{s}_\text{n}$ & $\skl{m}_\text{n}$ & $Q$\\
    & & $\EkinOverEpotf$ & $\EthermOverEpotf$ & [kyr] & & & $[M_\odot]$ & [pc$^{-3}$] & [$10^3$\,AU]&  &  & \\
    \hline                                 
    TH-m-1   &  3.3 & 0.075 & 0.047 & 48.0 & 0.96 & $311$    & $\phn0.86$ & $5.50\times10^4$ &  $6.51$& $0.42$ & $0.11$ & $0.26$ \\
    TH-m-2   &  3.6 & 0.090 & 0.047 & 45.5 & 0.91 & $429$    & $\phn0.74$ & $8.00\times10^4$ &  $8.51$& $0.65$ & $0.14$ & $0.21$ \\
    BE-c-1   &  3.3 & 0.058 & 0.039 & 27.5 & 0.55 & $305$    & $\phn0.94$ & $1.70\times10^6$ &  $3.11$& $0.16$ & $0.09$ & $0.53$ \\
    BE-c-2   &  3.6 & 0.073 & 0.039 & 27.5 & 0.55 & $331$    & $\phn0.97$ & $3.60\times10^4$ &  $5.68$& $0.31$ & $0.08$ & $0.27$ \\
    BE-m-1   &  3.3 & 0.053 & 0.039 & 30.1 & 0.60 & $195$    & $\phn1.42$ & $3.20\times10^6$ &  $1.10$& $0.13$ & $0.13$ & $1.03$ \\
    BE-m-2   &  3.6 & 0.074 & 0.039 & 31.9 & 0.64 & $302$    & $\phn0.54$ & $2.48\times10^6$ &  $1.46$& $0.13$ & $0.09$ & $0.74$ \\
    BE-s-1   &  3.3 & 0.055 & 0.039 & 30.9 & 0.62 & $234$    & $\phn1.14$ & $3.70\times10^7$ &  $0.52$& $0.11$ & $0.14$ & $1.30$ \\
    BE-s-2   &  3.5 & 0.074 & 0.039 & 35.9 & 0.72 & $325$    & $\phn0.51$ & $3.20\times10^6$ &  $1.43$& $0.21$ & $0.14$ & $0.68$ \\
    PL15-c-1 &  3.3 & 0.056 & 0.038 & 25.7 & 0.51 & $194$    & $\phn8.89$ & $2.42\times10^6$ &  $1.99$& $0.11$ & $0.06$ & $0.71$ \\
    PL15-c-2 &  3.6 & 0.068 & 0.038 & 25.8 & 0.52 & $161$    & $12.3\phn$ & $1.66\times10^4$ &  $7.82$& $0.45$ & $0.09$ & $0.21$ \\
    PL15-m-1 &  3.3 & 0.050 & 0.038 & 23.8 & 0.48 & $\phnn1$ & $20.0\phn$ & $-             $ &  $-$   & $-$    & $-$    & $-$    \\
    PL15-m-2 &  3.6 & 0.071 & 0.038 & 31.1 & 0.62 & $308$    & $\phn6.88$ & $2.66\times10^6$ &  $1.21$& $0.11$ & $0.11$ & $0.99$ \\
    PL15-s-1 &  3.3 & 0.053 & 0.038 & 24.9 & 0.50 & $\phnn1$ & $20.0\phn$ & $-             $ &  $-$   & $-$    & $-$    & $-$    \\
    PL15-s-2 &  3.5 & 0.069 & 0.038 & 36.0 & 0.72 & $422$    & $\phn4.50$ & $1.11\times10^7$ &  $1.01$& $0.16$ & $0.19$ & $1.20$ \\
    PL20-c-1 &  3.3 & 0.042 & 0.029 & 10.7 & 0.21 & $\phnn1$ & $20.0\phn$ & $-             $ &  $-$   & $-$    & $-$    & $-$    \\
    \hline                                 
  \end{tabular}
  
  \medskip
  The acronym for the run is shown in the first column, where the first part indicates the density profile, the middle letter the turbulent mode ('c' for compressive modes, 's' for solenoidal modes and 'm' for a natural mix of both), and the number at the end of each name the random seed for the turbulence. The initial energetic state is given by the Mach number $\mathcal{M}$, and the ratios of kinetic and thermal energy to the potential energy. $t_\mathrm{sim}$ and $t_\mathrm{sim}/t_\mathrm{ff}$ show the simulation time, $N_\mathrm{sink}$ the total number of protostars, and $M_\mathrm{max}$ the mass of the most massive protostar. The stellar number density is shown in column $n_*^\mathrm{glob}$. The global cluster properties are given as the mean separation between the protostars $\skl{s}$, the normalised mean separation $\skl{s}_\text{n}$, the normalised mean length of the minimal spanning tree $\skl{m}_\text{n}$, and the ratio $Q$.
\end{table*}

\section{Cloud and Cluster Analysis}
\label{sec:analysis}

In this section we briefly motivate and summarise the methods we used to analyse our simulation data.

\subsection{Energy Analysis}
The global energy partitioning of a gas cloud can be quantified by the ratio of kinetic to the potential energy $E_\mathrm{kin}/|E_\mathrm{pot}|$, where a value of $0.5$ corresponds to a virialised cloud. During the collapse of the cloud and the collapse of fragments into protostars, potential energy is converted into kinetic energy and transfered from the smooth gas to relatively compact protostars. In order to investigate the energy evolution of the collapse, we analyse the energy budget for the gas and the protostars separately. 

The total kinetic energy of the gas is calculated by simply summing over all cells in the cloud
\begin{equation}
  E_\mathrm{kin, gas} = \frac 1 2\,\sum_i m_i\,(v_{i, x}^2 + v_{i, y}^2 + v_{i, z}^2).
\end{equation}
The kinetic energy of the protostars, $E_\mathrm{kin,sink}$, is found analogously. For the potential energy of the gas we integrate numerically over radial bins around the centre of mass, yielding
\begin{equation}
  E_\mathrm{pot, gas}(r) = -\int\frac{G\,M(r)\mathrm{d}m(r)}{r},
\end{equation}
where $G$ is Newton's constant, $M(r)$ the enclosed mass inside radius $r$, and $\mathrm{d}m(r)$ the mass in the radial shell with thickness $\mathrm{d}r$. The potential energy of the protostars can be calculated by summing over the point masses
\begin{equation}
  \label{eq:Epot-summation}
  E_\mathrm{pot,sink} = -\sum_{i\neq j}G\,\frac{m_i\,m_j}{|r_i-r_j|}.
\end{equation}
However, in order to avoid the formation of hard binary systems and resulting very small time steps, we apply a softening term in the computation of the gravitational force between the protostars. For the softening we use the energy-conserving formalism described in \citet{Price07a} which yields a potential energy of
\begin{equation}
  \label{eq:Epot-softening}
  E_\mathrm{pot,sink} = \sum_{i\neq j}G\,m_i m_j\,\phi(r_i-r_j, h),
\end{equation}
with a kernel function $\phi$ (see Appendix~\ref{sec:smoothing}). On the one hand, the applied softening artificially prevents the formation of hard binaries and close orbits of particles in the simulation. On the other hand, it is questionable to what extend hard binaries can form in the early evolutionary phase. Numerically, the protostars are point objects with arbitrarily close separations. Physically, the protostars are very young and still in the contraction phase. Consequently, they have a relatively large sizes and low density contrasts in comparison to main sequence stars. Therefore, a dynamical treatment as extended gas spheres might well be more realistic. However, the detailed substructure inside the sink particle radius and the resulting dynamics is not captured in our simulations.

The internal motions of the gas and the protostars are quantified using the mass-weighted velocity dispersion
\begin{equation}
  \sigma_k^2 = \frac{\sum_im_i(u_{k,i}-\skl{u_k})^2}{\sum_im_i}
\end{equation}
where $k \in \{x,y,z\}$ and $\skl{u_k}$ is the mean velocity in dimension $k$,
\begin{equation}
  \skl{u_k} = \frac{\sum_im_i\,u_{k,i}}{\sum_im_i}.
\end{equation}
 The three-dimensional velocity dispersion is then given by
\begin{equation}
  \sigma_\mathrm{3D} = \sqrt{\sum_k\sigma_k^2}.
\end{equation}
In the simulations we calculate $\sigma_\mathrm{3D}$ using each component of the velocity. For the one-dimensional velocity dispersion we assume the same value for all three components and thus use $\sigma_\mathrm{1D}=\sigma_\mathrm{3D}/\sqrt{3}$. So far we have only considered the turbulent contribution to the velocity dispersion. Including the thermal contribution, the total dispersion along the line of sight is
\begin{equation}
  \sigma_\mathrm{tot} = \sqrt{\sigma_\mathrm{1D}^2 + c_\mathrm{s}^2\,}.
\end{equation}

\subsection{Subclustering}
Depending on the interplay between turbulent motions and the central collapse of a cloud, the spatial distribution of protostars may vary significantly (see Paper~I). In order to analyse the clustering properties of our protostars, we use the $Q$ value \citep{Cartwright04}
\begin{equation}
  \label{eq:Q-value}
  Q=\frac{\skl{s}_\mathrm{n}}{\skl{m}_\mathrm{n}}
\end{equation}
of the clusters. Here, $\skl{s}_\mathrm{n}$ is the normalised mean separation of the protostars and $\skl{m}_\mathrm{n}$ is the normalised mean length of the edges of the minimal spanning tree (MST), where the edge is the distance between two protostars. For a detailed discussion of the motivation for this definition of $Q$ see \citet{Cartwright04}.

The distribution function $p(s)$ describes the probability of two protostars to be separated by the distance, $s$. We discretise $p(s)$ with $N_\mathrm{bin}$ bins for the entire cluster, leading to an equal-sized bin width of $\Delta s = 2R_\text{C}/N_\text{bin}$, where $R_\text{C}$ is the cluster radius. The normalised number of pairs in bin $i$ can thus be expressed with
\begin{equation}
  \label{eq:separation-distribution-function}
  p(i) = \frac{2N_i}{N_\text{C}(N_\text{C}-1)\Delta s}.
\end{equation}
Here $N_i$ denotes the number of pairs with a distance in the range $[i\Delta s,(i+1)\Delta s)$ and $N_\text{C}(N_\text{C}-1)/2$ is the total number of separations for $N_\text{C}$ cluster members. Multiple peaks in the distribution function are related to subcluster structure, which gives higher counts at low distances due to the small separations within each subcluster and higher counts at a larger separation due to the large distance between the subclusters. In case of no distance degeneracy between subclusters, the number of peaks equals the number of subclusters. The mean value $\skl{s}$ of all $N_\text{C}(N_\text{C}-1)$ particle separations $s_j$,
\begin{equation}
  \skl{s} = \frac{2}{N_\text{C}(N_\text{C}-1)}\sum_j s_j,
\end{equation}
gives a measure for the mean distance between particles in the set.

The MST is calculated using the \citet{GowerRoss69} description of Prim's algorithm \citep{Prim57}. The more particles are confined in an observed area, the smaller is the mean edge of the tree. The resulting decrease of the mean edge due to the increasing number of nodes in the tree has to be corrected by a dimensionality factor. The correction factor for the three-dimensional cluster model with cluster volume $V$ was set to
\begin{equation}
  \frac{(VN_\text{C}^2)^{1/3}}{N_\text{C}-1},
\end{equation}
taken from \citet{Schmeja06}.

For stellar clusters with a smooth radial density gradient, $Q$ ranges from $0.8-1.5$, corresponding to a radial density distribution of particles $n\propto r^{-\eta}$ with $\eta=0$ to $2.9$. Clusters with substructure have $Q=0.8-0.45$, decreasing with increasing degree of subclustering. A detailed relation between $Q$, $\eta$ and the degree of subclustering can be found in \citet{Cartwright04}.

\subsection{Mass Segregation}
A set of stars or protostellar objects may show a mass-dependent spatial distribution within a cluster. In a mass-segregated cluster, massive objects tend to be located closer to the centre of the cluster, whilst low-mass objects occupy regions of larger radii. We quantify the degree of mass segregation using the MST as described in \citet{Allison09} with the \emph{mass segregation ratio} (MSR)
\begin{equation}
  \label{eq:mass-segregation-ratio}
  \Lambda_\mathrm{MSR} = \frac{\skl{l_\mathrm{norm}}}{l_\mathrm{massive}}\pm\frac{\sigma_\mathrm{norm}}{l_\mathrm{massive}}.
\end{equation}
The ratio describes, how large the spatial spread of the most massive stars is, compared to the spatial spread of a random choice of stars. How many most massive stars are counted and compared to an equal amount of random stars should not be fixed, but rather treated as a free parameter, which we name $N_\mathrm{MST}$. In order for the MST of the random set of stars to be a good measure for the average spread, we need to pick many sets of random stars and average over the individual lengths of the MST. We set the number of sets to $500$ as suggested by \citet{Allison09}. With the average length $\skl{l_\mathrm{norm}}$ of these $500$ sets and the length of the $N_\mathrm{MST}$ most massive stars, $l_\mathrm{massive}$, we then determine the degree of mass segregation. The error is computed with the standard deviation $\sigma_\mathrm{norm}$ of $\skl{l_\mathrm{norm}}$.
If $\Lambda_\mathrm{MSR}$ takes values significantly larger than unity, the $N_\mathrm{MST}$ most massive stars are located much closer to one another than the same amount of randomly picked stars. Hence the system shows mass segregation. In the opposite case ($\Lambda_\mathrm{MSR}\ll1$) the most massive stars have much larger distances between one another than a set of random stars in the cluster and the system shows inverse mass segregation. $N_\mathrm{MST}$ is basically a free parameter that we loop over starting from 2 up to half of the total number of sink particles, in order to determine the number $N_\mathrm{MST}$ up to which the system is mass segregated, i.e., $\Lambda_\mathrm{MSR}>1$. The weak point of this method is its sensitivity to massive outlier protostars, in particular protostars among the $N_\mathrm{MST}$ most massive stars that are clearly located between subclusters in an environment of many separate clusters \citep[see,][]{MaschbergerClarke2011}. As we do not investigate the total cloud with this method but only reduced clusters without outliers (see section~\ref{sec:reduced-cluster}) our results are not affected by this behaviour.

Mass segregation can either originate from dynamical $N$-body relaxation or is primordial in nature, where the latter case means the more massive stars form closer to the centre. In order to analyse whether mass segregation is primordial or due to dynamical processes, we use the mass segregation time \citep{Spitzer69},
\begin{equation}
  t_\mathrm{seg}(M) \approx \frac{\skl{m}}{M}\,t_\mathrm{relax},
\end{equation}
with $\skl{m}$ being the average mass of all stars in the cluster and $M$ the mass of the star in question.
The relaxation time $t_\mathrm{relax}$ can be expressed in terms of the number of stars $N$, the radius of the cluster $R_\mathrm{C}$, and the stellar velocity dispersion $\sigma$, yielding for the mass segregation time \citep[e.g.][]{BinneyTremaine87},
\begin{equation}
  \label{eq:mass-segregation-time}
  t_\mathrm{seg}(M) \approx \frac{\skl{m}}{M}\,\frac{N}{8\ln N}\,\frac{R_\mathrm{C}}{\sigma}.
\end{equation}
By setting the time according to different stages in the simulation, one can obtain the minimum mass down to which stars had enough time to dynamically mass segregate. Care must be taken when applying the mass segregation time to hydrodynamic collapse simulations. In contrast to old stellar clusters, where there is no or very little interstellar gas left and consequently $N$, $\skl{m}$ and $M$ do not vary with time, hydrodynamic collapse simulations follow the formation of protostars from the beginning of the collapse. Not only do protostars form at different times, they also accrete further gas from the surrounding dense medium in which they were born and are subject to gas drag forces. The number of protostellar objects $N$, their individual masses $M$, their mean mass $\skl{m}$, and the cluster radius $R$ are therefore strongly varying with time. Consequently, the mass segregation and the minimum segregated mass for a given time can not be calculated for the total set of objects as a whole. Instead, the possibility of being segregated within the cluster has to be estimated for each star individually by taking into account the formation time and the growing mass of the star due to accretion.

\section{Results}
\label{sec:results}

\subsection{Overview}
We follow the cloud collapse until 20\% of the mass is accreted by sink particles. The simulation time, the number of formed protostars, the mass of the most massive protostar and the key parameters of the substructure of the cluster are listed in table~\ref{tab:simulation-global-overview}. A column density plot at the end of each simulation is shown in Paper~I, figures~4 and 5.

The TH profile takes the longest time to form gravitationally collapsing regions and to capture $20~M_\odot$ in sink particles. During this time, approximately $45-50~\mathrm{kyr}$, the turbulent motions can compress the gas in locally disconnected areas, leading to distinct subclusters of sink particles. The stronger mass concentration in the centre of the BE setups and the resulting shorter collapse and sink particle formation time suppresses the formation of disconnected subclusters in favour of one main central cluster (see morphology in Paper~I). The corresponding PL15 profiles show a very similar overall cloud structure to the BE runs, but significantly different stellar properties. Due to the much stronger gas concentration in the centre of the cloud, all PL15 setups form a protostar very early in the simulation. This initial central protostar accretes the surrounding gas at a high rate and can grow to a massive protostar before the turbulent motions eventually form collapsing filaments and trigger fragmentation. The PL15 setups with turbulent fields m-1 and s-1 (PL15-m-1, PL15-s-1) do form dense filaments, but no further sink particles until the first protostar reaches a mass of $20~M_\odot$. In case of multiple sink particles, the clusters are more compact than in the corresponding BE case. The PL20 profile only forms one single sink particle due to the very strong mass concentration. The central protostar forms very early and accretes gas at an almost constant rate of $\approx2\times10^{-3}~M_\odot~\mathrm{yr}^{-1}$, close to the analytical value of a highly unstable singular isothermal sphere (\citealt{Shu77}; Paper~I). This results in a total simulation time of only $11~\mathrm{kyr}$, which is not enough for turbulent motions to form filaments and further sink particles.

The following discussion of the cluster properties therefore abstains from a detailed description of the setups PL15-m-1, PL15-s-1 and PL20-c-1.

\subsection{Energy Evolution of the Global Cloud}
In order to better understand the energy evolution, we separately analyse the gas and sink particle contributions to the total energy. 

\begin{figure}
  \centering
  \includegraphics[width=8cm]{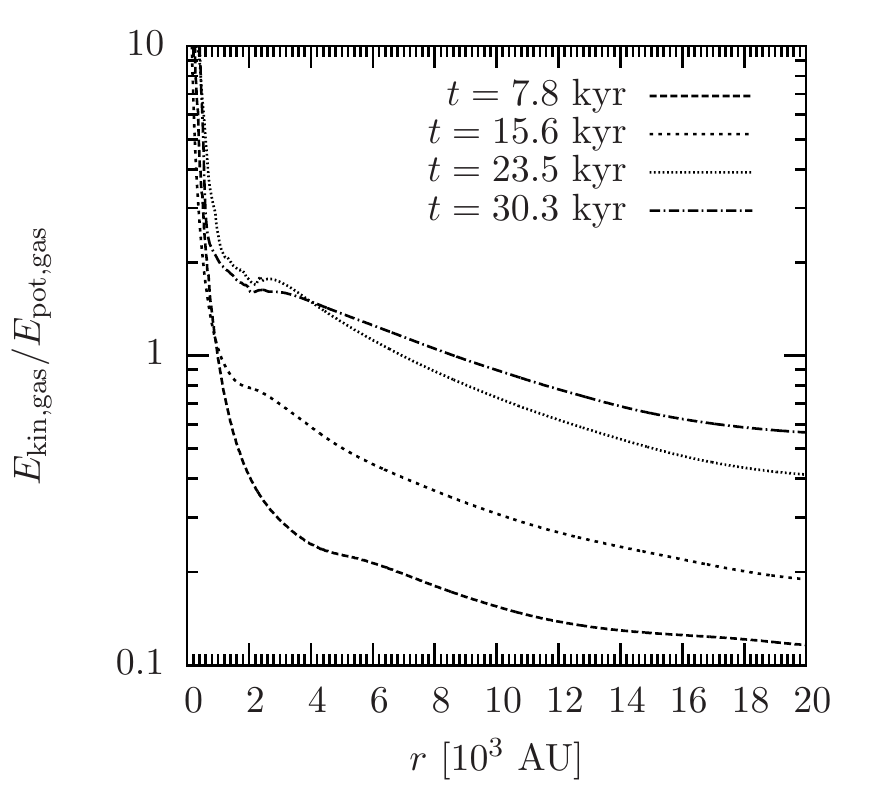}
  \caption{Ratio of kinetic over potential energy of the gas for the BE-m-1 setup as a function of radius for different times. The cloud evolves from a strongly gravitationally dominated state to an energy state with $E_\text{kin,gas}/|E_\text{pot,gas}|>0.5$ at the end of the simulation.}
  \label{fig:BE-mix-1-EkinEpot}
\end{figure}

All setups are gravitationally very unstable and start to collapse immediately. As a result, the initial random velocities of the gas are reoriented towards the direction of the central acceleration. The total kinetic energy strongly increases with time due to the infall motion. Figure~\ref{fig:BE-mix-1-EkinEpot} shows a representative example of the kinetic over the potential energy of the gas as a function of radius for different times in the simulation. The cloud starts in a strongly sub-virial state and exceeds a ratio of kinetic to gravitational energy of 0.5 for the entire cloud after roughly $20~\mathrm{kyr}$. Within a radius of $10^4~\mathrm{AU}$ the ratio reaches values greater than unity and diverges in the very central region. This behaviour can be explained by a simple estimate using a singular isothermal sphere, which is characterised by an initial density profile $\rho\propto r^{-2}$ and approaches a free-fall density profile $\rho\propto r^{-3/2}$ inside the head of the rarefaction wave \citep{Shu77,WhitworthSummers1985}. The corresponding velocity field scales as $v\propto r^{-1/2}$. The resulting potential energy scales as $E_\mathrm{pot}\propto r^2$, while the kinetic energy follows a relation $E_\mathrm{kin}\propto r^{1/2}$. Consequently, the ratio $E_\mathrm{kin}/|E_\mathrm{pot}|$ scales as $r^{-3/2}$ and diverges for small radii, indicating that the innermost part of the cloud is dominated by kinetic energy.

\begin{figure}
  \centering
  \includegraphics[width=8cm]{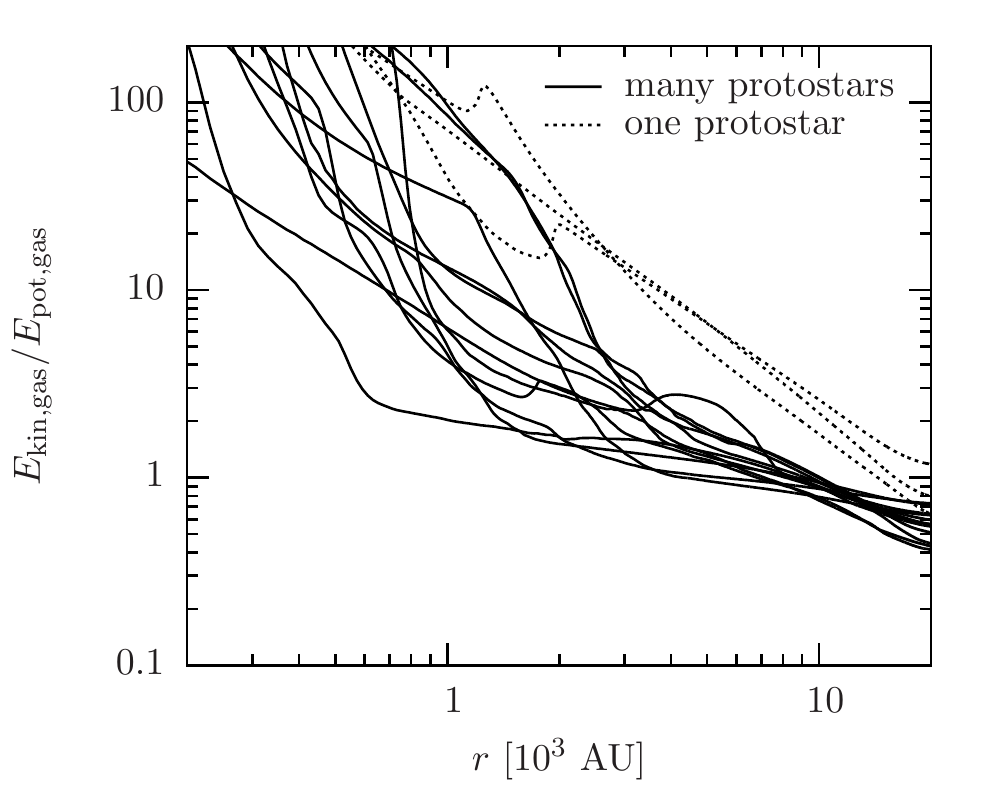}
  \caption{Ratio of kinetic over potential energy of the gas for all profiles at the end of the simulation at 20\% star formation efficiency. The dotted lines indicate the runs with only one protostar, the runs with multiple protostars are shown with solid lines. Note that the physical times differ strongly between $11$ and $48$ kyr for the different setups, see table~\ref{tab:simulation-global-overview}.}
  \label{fig:EkinEpot-gg-end}
\end{figure}

The different initial density profiles as well as the different formation modes of protostars lead to different radial distributions during the collapse. A comparison of $E_\text{kin,gas}/|E_\text{pot,gas}|$ for all setups at the end of the simulation is shown in figure~\ref{fig:EkinEpot-gg-end}. A significant difference is found between the simulations with only one protostar (dotted lines) and the ones that form many protostars (solid lines). The three setups with only one protostar show much higher values for most of the cloud and a steeper slope. This is not surprising because the gas in the central region can fall towards the central particle without being disturbed by other sink particles and their $N$-body interactions. In case of multiple protostars the ratio $E_\text{kin,gas}/|E_\text{pot,gas}|$ shows a large scatter close to the central region ($R\lesssim4\times10^3~\mathrm{AU}$), which can be explained by the local variations in the sink particle positions and motions, and the resulting impact on the gas. The scatter in the energy ratio is significantly lower in the outskirts of the cluster.

\begin{figure}
  \centering
  \includegraphics[width=8cm]{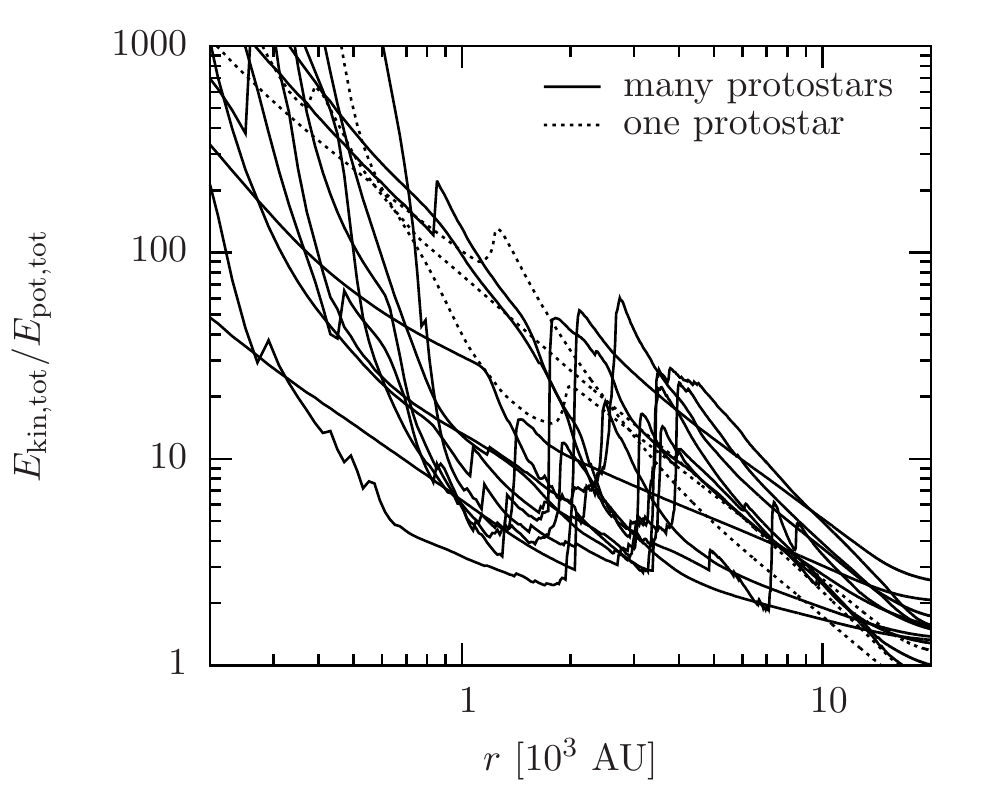}
  \caption{Same as figure~\ref{fig:EkinEpot-gg-end} but for the total energies (protostars and gas).}
  \label{fig:EkinEpot-tt-end}
\end{figure}

The average value as well as the spread of $E_\text{kin,tot}/|E_\text{pot,tot}|$ increase when the sink particles' mass is included in the virial analysis (see figure~\ref{fig:EkinEpot-tt-end}).
There is no systematic correlation between the various initial conditions and the ratio of the energies. The fact that including the protostars leads to higher values, shows that the cluster contributes more to the kinetic rather than the potential energy. At this point, we want to emphasise that the computation of the potential energy with and without gravitational softening shows different values that vary by a factor of a few. This also influences the kinetic energy evolution of protostars in close encounters. Considering the large values of kinetic over potential energy and the large spread in the central regions of the clouds, this does not affect the overall result that the central region is strongly dominated by kinetic energy. On longer evolutionary time scales, one has to take into account that the very young protostars are still large objects that slowly contract. The effects of protostellar collisions at different stages, the resulting mergers and formed binaries may have different effects on the global energy evolution.

\begin{figure}
  \centering
  \includegraphics[width=8cm]{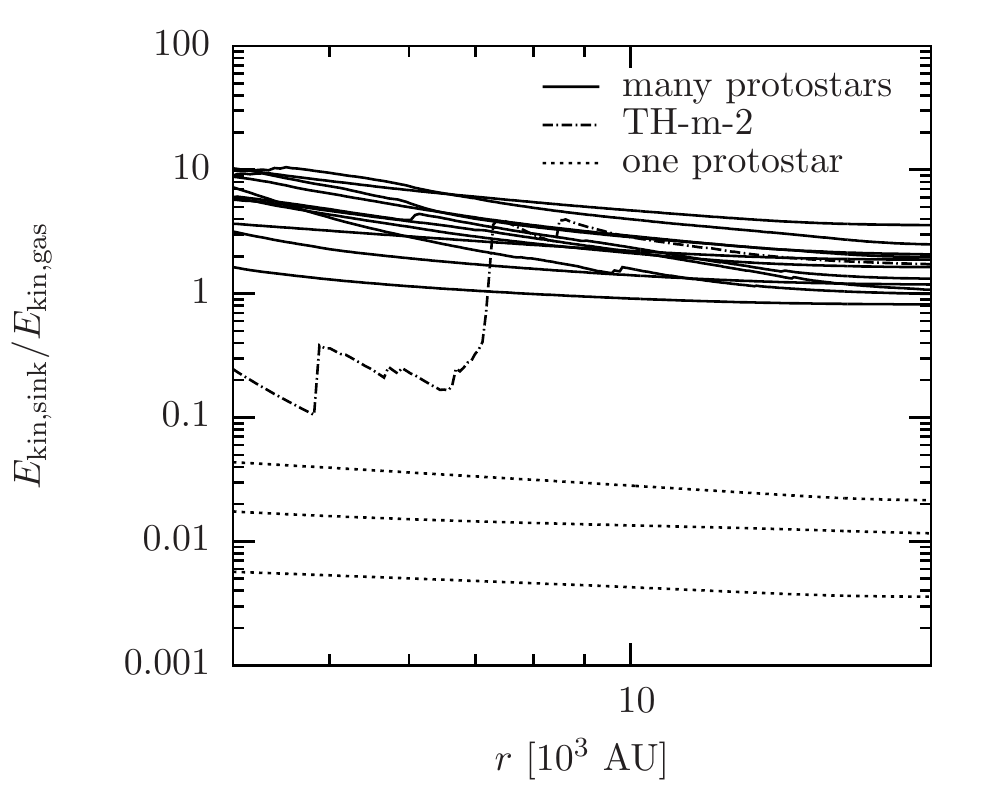}
  \caption{Ratio of kinetic energy of the sink particles to the kinetic energy of the gas for all profiles at the end of the simulation. The inner region ($R_\mathrm{C}<4~\mathrm{AU}$) varies extremely because of the slight offsets of the centre of mass and the centre of the cluster and is not shown.}
  \label{fig:EkinEkin-sg-end}
\end{figure}

A comparison of the kinetic energy of the sink particles and the gas ($E_\text{kin,sink}/E_\text{kin,gas}$) is plotted in figure~\ref{fig:EkinEkin-sg-end}. The ratio is above unity for all simulations with many protostars (solid lines). Although the protostars account for only 20\% of the total mass at the end of the simulation, their kinetic energy dominates the total kinetic energy budget of the cloud. Again, the setups with only one protostar constitute an exception (dotted lines). In these cases, the kinetic energy of the protostar is significantly lower, which can be explained by accretion flows from opposite directions that result in an almost vanishing net momentum transfer onto the protostar (see figure~\ref{fig:EkinEkin-sg-end}). The dashed-dotted line shows $E_\text{kin,sink}/E_\text{kin,gas}$ for the TH-m-2 setup. As the cloud in this run forms two distinct subclusters with a central void between them (see right part of figure~\ref{fig:TH-subclusters}) the total kinetic energy of the few protostars between the subclusters is relatively low.


\begin{figure}
  \centering
  \includegraphics[width=8cm]{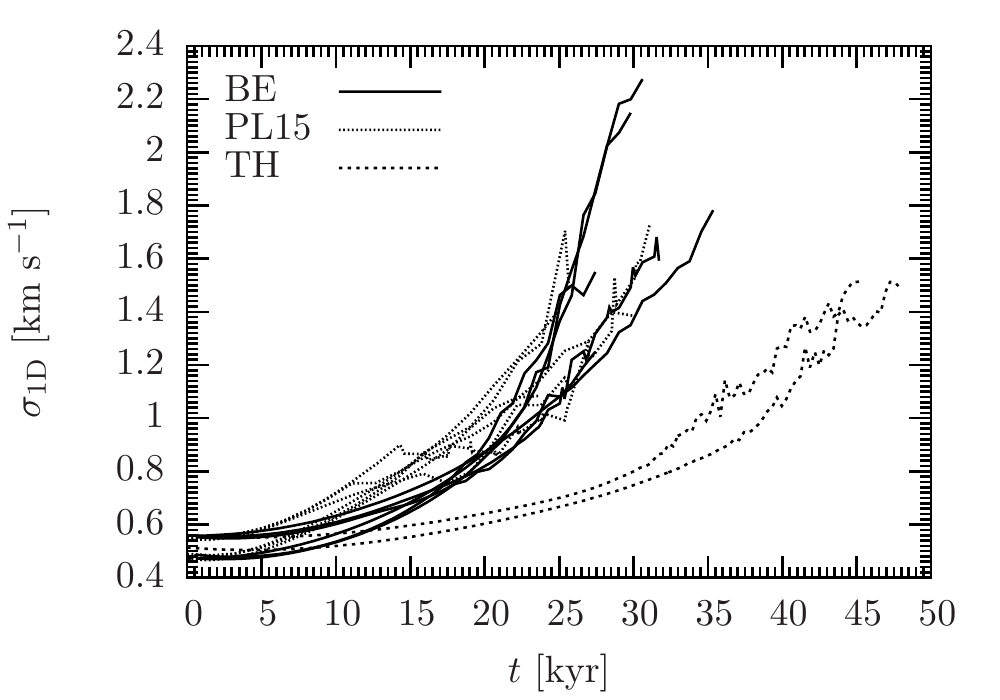}
  \caption{Velocity dispersion for the gas for all runs. The values increase over time due to the increasing infall motion.}
  \label{fig:velocity-dispersion-gas}
\end{figure}

\begin{figure}
  \centering
  \includegraphics[width=8cm]{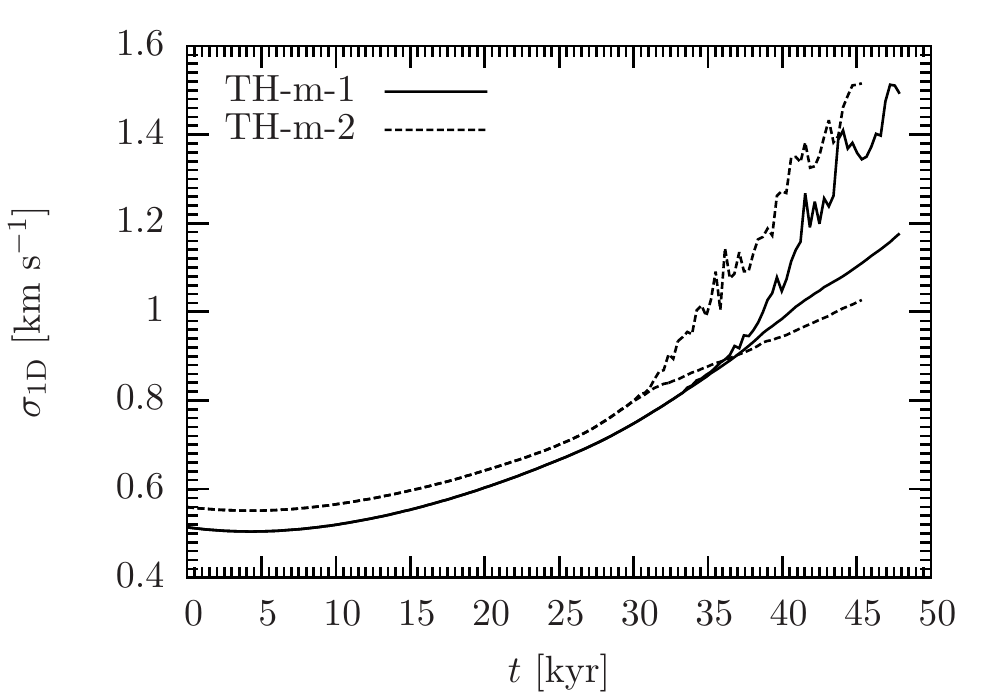}
  \caption{Velocity dispersion for the TH setups. The lower curves correspond to the gas only, the higher curves include the turbulent contribution of the sink particles. As the sink particles contain a significant fraction of the kinetic energy, the curves including sink particles are remarkably higher.}
  \label{fig:velocity-dispersion-sg-TH}
\end{figure}

As a link to observable properties of star-forming regions we calculate the velocity dispersion for the entire cloud as a function of time. Here we assume isotropy of the motions of the gas and restrict our analysis to the one-dimensional velocity dispersion $\sigma_\mathrm{1D}$. Because of the initial random turbulence, the velocity dispersion of the gas shows anisotropies, which tend to reduce during the simulation. Initially, the deviation from isotropy is of the order of $10-20\%$. During the simulation the value shows variations but decreases to about half of the initial value ($\Delta\sigma/\sigma\sim5-10\%$), averaged over all simulations. There is no clear trend with the varied initial conditions and the number of protostars. Figure~\ref{fig:velocity-dispersion-gas} shows the turbulent velocity dispersion $\sigma_\mathrm{1D}$ for the gas for all runs. Initially, $\sigma$ just reflects the initial turbulent velocity, the increasing values correspond to the additional infall motion. The significantly lower values for the TH profiles are simply due to the delayed dominant central collapse. The formation of disconnected subclusters reduces the global infall speed in comparison to the other setups with one central cluster. The combined velocity dispersion for gas and sink particles can be seen for the TH profiles in figure~\ref{fig:velocity-dispersion-sg-TH}. The plots for the other setups look similar. As shown in figure~\ref{fig:EkinEkin-sg-end}, the protostars contain a significant fraction of the kinetic energy. Therefore, the total value including sink particles is remarkably higher. None of the curves saturates during the simulated time, which can be explained by a simple free-fall approximation. The maximum speed that can be reached by free-falling gas is of the order of $R_0/t_\mathrm{ff}\approx2~\mathrm{km~s}^{-1}$, where $R_0$ is the cloud radius and $t_\mathrm{ff}$ the global free-fall time. None of the setups needs more than a free-fall time to convert $20\%$ of the gas mass into stars when we stop the simulation, so no setup had enough time to reach the limiting free-fall velocity dispersion of $2~\mathrm{km~s}^{-1}$.

\begin{figure}
  \centering
  \includegraphics[width=8cm]{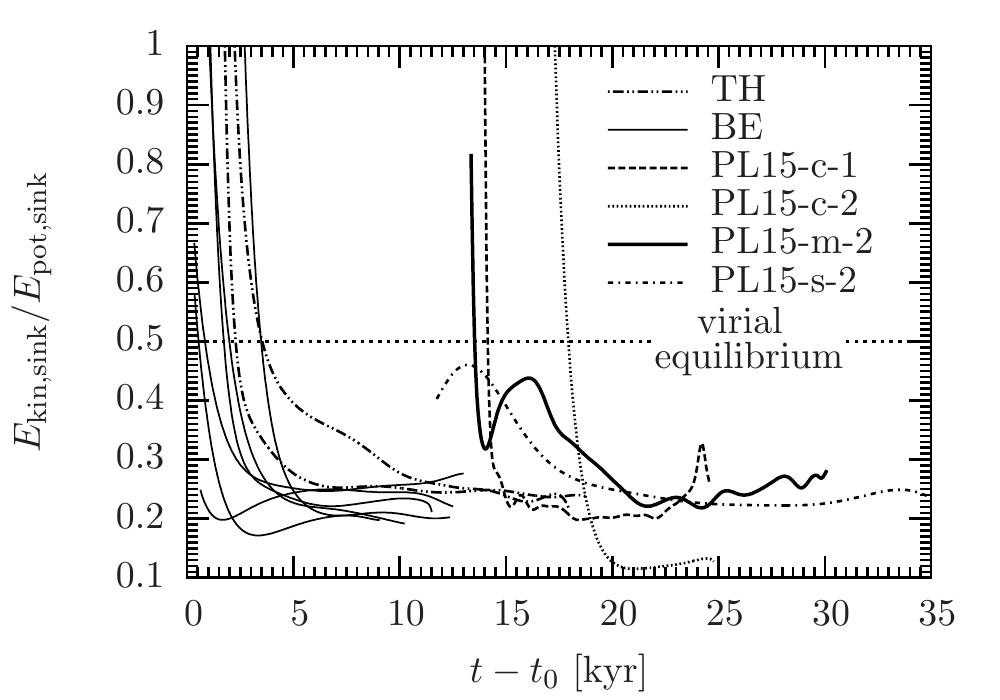}
  \caption{Ratio of kinetic to potential energy as a function of time for the sink particles only. The time was adjusted to the formation of the first sink particle in the setup.}
  \label{fig:EkinEpot-sp}
\end{figure}

With a focus on the nascent cluster as an $N$-body system, we also analyse the virial state of the sink particles without including the contributions of the surrounding gas. To do so, we treat the protostars as point masses and we calculate the gravitational potential via direct summation (equation~\ref{eq:Epot-softening}). The corresponding ratio of kinetic to potential energy for the sink particles is shown in figure~\ref{fig:EkinEpot-sp}, excluding the runs with only one protostar. The time axis in the plot is adjusted to the time when the first condensation was created. In the case of all PL15 profiles with multiple sink particles, the second and further sink particles formed with a large delay after the first sink particle. Therefore, the curves for the PL15 profiles start at times $t-t_0 > 10~\mathrm{kyr}$ (see $\Delta t_{12}$ in table~\ref{tab:reduced-cluster-properties}).
The first protostars form with the velocity that the collapsing condensation inherits from the gas motion. The positions at which they form is determined by the structure of the filaments or the fragmenting disc. As they form independent from one another, sometimes even in separate subclusters, their kinetic and potential energies are uncorrelated. As the protostars form with the velocity of the parental gas cloud and because they are usually separated by a large distance, the initial values of $E_\text{kin,sink}/|E_\text{pot,sink}|$ are very high. Soon after their formation, the protostars dynamically decouple from the gas and move towards the central region of the nascent cluster. The system begins to virialise, leading to decreasing values of $E_\text{kin,sink}/|E_\text{pot,sink}|$. Without the formation of subsequent protostars, the system would quickly reach a virialised state. However, as this process continues, the energy ratio of the total cluster is influenced by the virial state of the newly formed objects. If they form at time $t_i$ at position $r_i$ with velocities $v_i$ smaller than the virial velocity $v_\mathrm{virial}(r_i,t_i)$, they lead to a decreasing energy ratio. A quick analytical estimate illustrates, why this behaviour is expected. The virial velocity is given by
\begin{equation}
  v_\mathrm{virial} = \rkl{\frac{GM_\mathrm{Cl}}{R_\mathrm{Cl}}}^{1/2},
\end{equation}
with the mass and radius of the cluster $M_\mathrm{Cl}$ and $R_\mathrm{Cl}$. As a lower limit, we can assume a constant stellar density in the cluster over time, $\rho_*$, which relates the cluster radius to the cluster mass like $R_\mathrm{Cl}(t) = \rkl{3M_\mathrm{Cl}(t)/(4\pi\rho_*)}^{1/3}$ and thus the virial velocity in this lower limit follows $v_\mathrm{virial,low}\propto M_\mathrm{Cl}^{1/3}$, increasing with time as the total mass of the cluster increases. Of course, the velocity of the gas is also increasing over time due to the collapse of the cloud. However, as shown in figure~\ref{fig:velocity-dispersion-gas}, the velocity dispersion of the gas increases over time by a factor of only 3 at most. In addition, figure~\ref{fig:velocity-dispersion-sg-TH} illustrates that the kinetic contribution of the protostars is remarkably larger than that of the gas. In order for the lower limit virial velocity, $v_\mathrm{virial,low}$, to be higher than the average gas velocity, the cluster mass must grow by a factor of 27 during the entire simulation, which can be achieved. Considering the fact, that the stellar density also increases, the virial velocity will be even higher. Consequently, the newly formed stars, which inherit the low gas velocity, tend to decrease the energy ratio.

\begin{figure}
  \includegraphics[width=8cm]{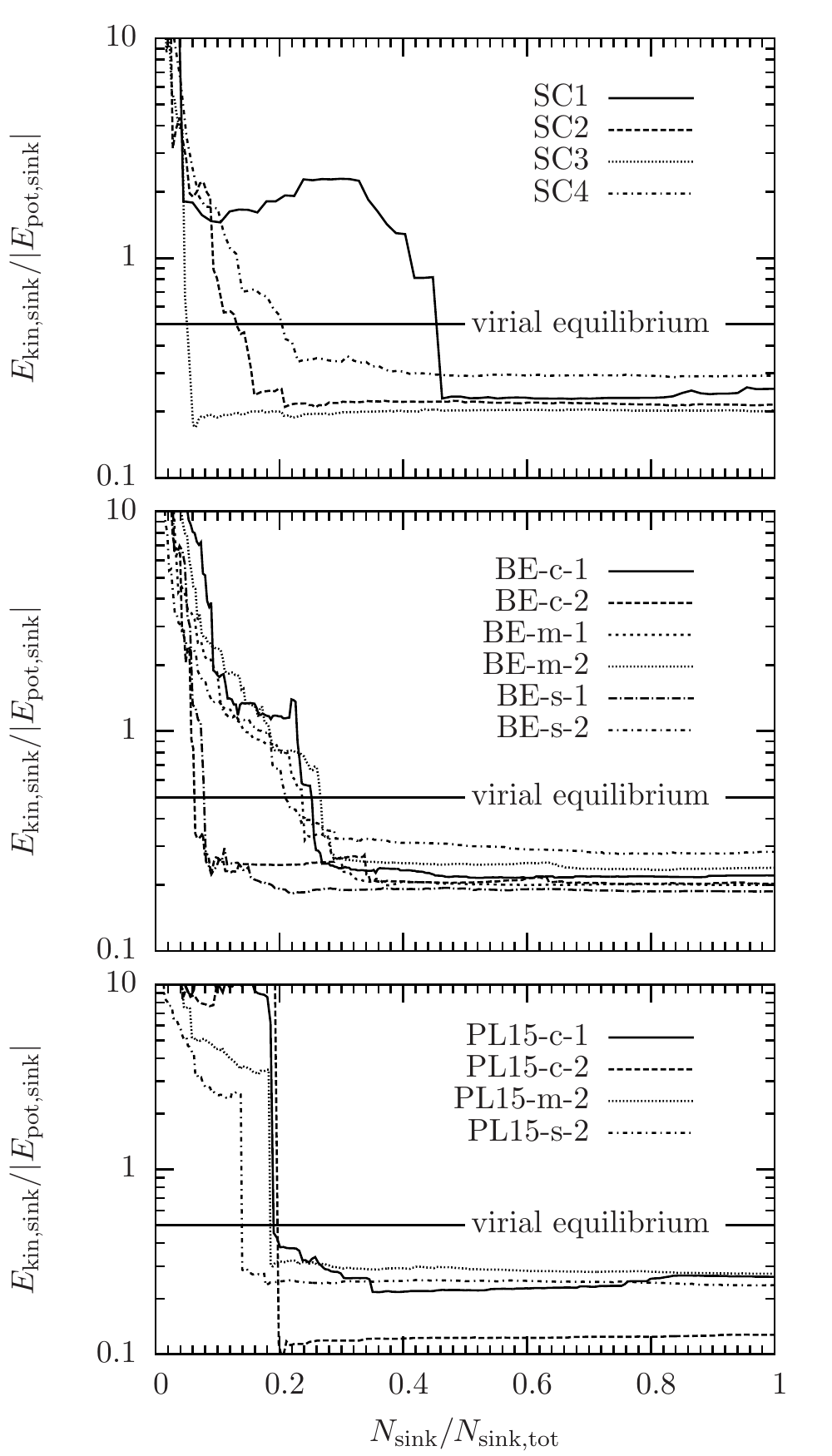}
  \caption{Energy balance $E_\text{kin,sink}/|E_\text{pot,sink}|$ as a function of normalised number of protostars, $N_\mathrm{sink}/N_\mathrm{sink,tot}$. The protostars were sorted by their distance from the centre of the cluster. In all cases, only the innermost $\sim10-30\%$ of the stars form a cluster with virial or super-virial energy balance.}
  \label{fig:radial-virial-state}
\end{figure}

The larger the cluster, the lower is the available mass in the central region of the cluster \citep[see also,][]{Girichidis11b}. Therefore, new protostars must form at increasingly larger radii. In order to show that these new stars are the ones that push the ratio $E_\text{kin,sink}/|E_\text{pot,sink}|$ to lower than virialised values, we calculate the ratio as a function of the fraction of total protostars. Figure~\ref{fig:radial-virial-state} shows the energy ratio with the protostars sorted by their distance from the centre of the cluster. In all cases, only the innermost $\sim10-30\%$ have a virial or super-virial energy balance. The majority of the nascent cluster has an overall sub-virial energy partition. But we expect that the ensemble virialises on a dynamical time scale as soon as star formation stops in the cluster region.

From our simulations we conclude that a detailed energy analysis can only be performed properly, if both protostars and gas are included in a self-consistent way. In turn, the remaining gas is essential to the virial state of the nascent cluster.

\subsection{Global Cluster Properties}

In this section we discuss the spatial distribution of the protostars in the simulated cloud. We begin with an analysis of all protostars in the simulations in order to measure the cluster properties of the cloud as a whole. A detailed investigation of individual subclusters without outlier protostars is presented in section~\ref{sec:reduced-cluster} and below.

\begin{table*}
  \begin{minipage}{\textwidth}
    \caption{Reduced cluster properties for the simulations with many sink particles}
    \label{tab:reduced-cluster-properties}
    \begin{tabular}{lcccccccccccccc}
      Cluster&$N_\mathrm{sink}$&$M_\mathrm{C}$&$\skl{M}$&$R_\mathrm{C}$&$n_*^\mathrm{red}$&$t_1$&$\Delta t_{12}$&$\sigma_\mathrm{1D}$&$t_\mathrm{relax}$&$t_\mathrm{avail}$&$N_\mathrm{seg}$&$f_\mathrm{seg}$&seg?\\
             &      &$[M_\odot]$&$[M_\odot]$&[kAU]&[pc$^{-3}$]&[kyr]&[kyr]&[km/s]&[kyr]&[$t_\mathrm{relax}$]&&\\
      \hline
      SC1 (TH-m-1)& \phn67 & \phn4.2 & 0.063 & 2.74 & $6.82\times10^{6\phn}$ &    32.0 & \phn0.008 & 1.86 & \phn8.01 &\phn2.00 &  \phn19 & 0.28  & 0   \\
      SC2 (TH-m-1)& 182    & 10.4    & 0.057 & 0.97 & $4.18\times10^{8\phn}$ &    29.9 & \phn0.776 & 2.90 & \phn4.02 &\phn4.31 &  \phn72 & 0.40  & ++  \\
      SC3 (TH-m-2)& 232    & \phn9.4 & 0.041 & 1.00 & $4.86\times10^{8\phn}$ &    26.5 & \phn0.709 & 2.82 & \phn5.17 &\phn3.53 &  \phn82 & 0.35  & +   \\
      SC4 (TH-m-2)& 100    & \phn5.7 & 0.057 & 0.45 & $2.30\times10^{9\phn}$ &    28.5 & \phn0.933 & 3.00 & \phn1.12 &   14.31 &  \phn77 & 0.77  & ++  \\
      BE-c-1      & 192    & 11.4    & 0.060 & 0.64 & $1.53\times10^{9\phn}$ &    14.9 & \phn0.279 & 3.61 & \phn2.21 &\phn5.58 &  \phn81 & 0.42  & 0   \\
      BE-c-2      & 275    & 15.0    & 0.055 & 5.05 & $4.47\times10^{6\phn}$ &    15.1 & \phn0.764 & 2.43 & 34.93    &\phn0.33 &  \phnn8 & 0.03  & 0   \\
      BE-m-1      & \phn99 & 11.9    & 0.121 & 0.24 & $1.50\times10^{10}$    &    19.6 & \phn0.052 & 5.91 & \phn0.30 &   34.66 &  \phn84 & 0.85  & ++  \\
      BE-m-2      & 255    & 15.7    & 0.061 & 1.39 & $1.99\times10^{8\phn}$ &    20.2 & \phn0.086 & 3.20 & \phn6.81 &\phn1.71 &  \phn67 & 0.26  & +   \\
      BE-s-1      & 190    & 16.1    & 0.085 & 0.50 & $3.18\times10^{9\phn}$ &    21.5 & \phn0.083 & 4.64 & \phn1.35 &\phn6.92 &  100    & 0.53  & ++  \\
      BE-s-2      & 288    & 16.7    & 0.058 & 2.12 & $6.33\times10^{7\phn}$ &    22.3 & \phn0.004 & 2.85 & 12.97    &\phn1.05 &  \phn43 & 0.15  & $-$ \\
      PL15-c-1    & 170    & 17.0    & 0.100 & 1.46 & $1.14\times10^{8\phn}$ & \phn1.1 & 13.5\phnn & 4.02 & \phn4.11 &\phn2.69 &  \phn37 & 0.22  & ++  \\
      PL15-c-2    & \phn79 & 14.8    & 0.187 & 1.64 & $3.75\times10^{7\phn}$ & \phn1.0 & 15.5\phnn & 2.50 & \phn1.45 &\phn6.43 &  \phn13 & 0.16  & 0   \\
      PL15-m-2    & 240    & 15.6    & 0.065 & 1.00 & $5.03\times10^{8\phn}$ & \phn1.0 & 13.3\phnn & 4.47 & \phn3.34 &\phn5.03 &  \phn68 & 0.28  & $-$ \\
      PL15-s-2    & 396    & 18.5    & 0.047 & 1.46 & $2.67\times10^{8\phn}$ & \phn0.9 & 10.3\phnn & 3.45 & \phn9.64 &\phn2.57 &  \phn82 & 0.21  & 0   \\
      \hline
    \end{tabular}
    
    \medskip
    The table shows the properties for the reduced cluster with the number of protostars $N_\mathrm{sink}$, the total cluster mass $M_\mathrm{C}$, the average protostellar mass $\skl{M}$, the radius $R_\mathrm{C}$, and the protostellar number density $n_*^\mathrm{red}$. Column $t_1$ indicates the time of the formation of the first protostar, $\Delta t_{12}$ the time difference between the formation of the first and the second protostar. $\sigma_\mathrm{3D}$ and $\sigma_\mathrm{1D}$ show the stellar velocity dispersion of the cluster. The key values for the mass segregation are the relaxation time $t_\mathrm{relax}$, the available lifetime of the cluster $t_\mathrm{avail}$ in units of the relaxation time, and the total and normalised number of protostars that had enough time to relax dynamically $N_\mathrm{seg}$ and $f_\mathrm{seg} = N_\mathrm{seg}/N_\mathrm{sink}$. The column ``seg?'' indicates the segregation state of the cluster: significantly mass segregated (++), marginally mass segregated (+), not mass segregated (0), and inversely mass segregated ($-$).
  \end{minipage}
\end{table*}

\subsubsection{TH runs}
\begin{figure*}
  \begin{minipage}{\textwidth}
    \centering
    \includegraphics[width=8cm]{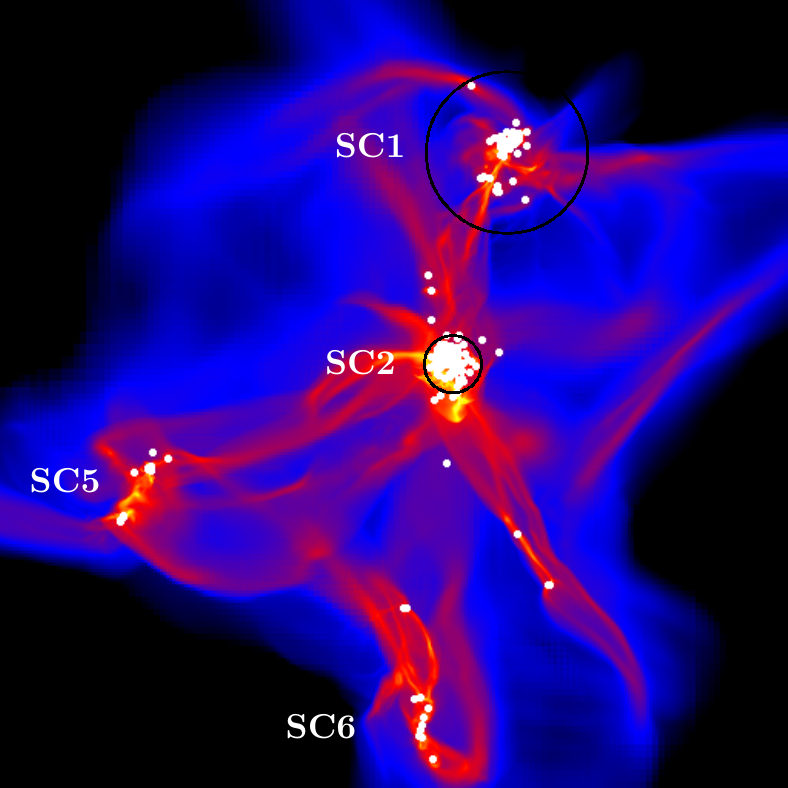}
    \includegraphics[width=8cm]{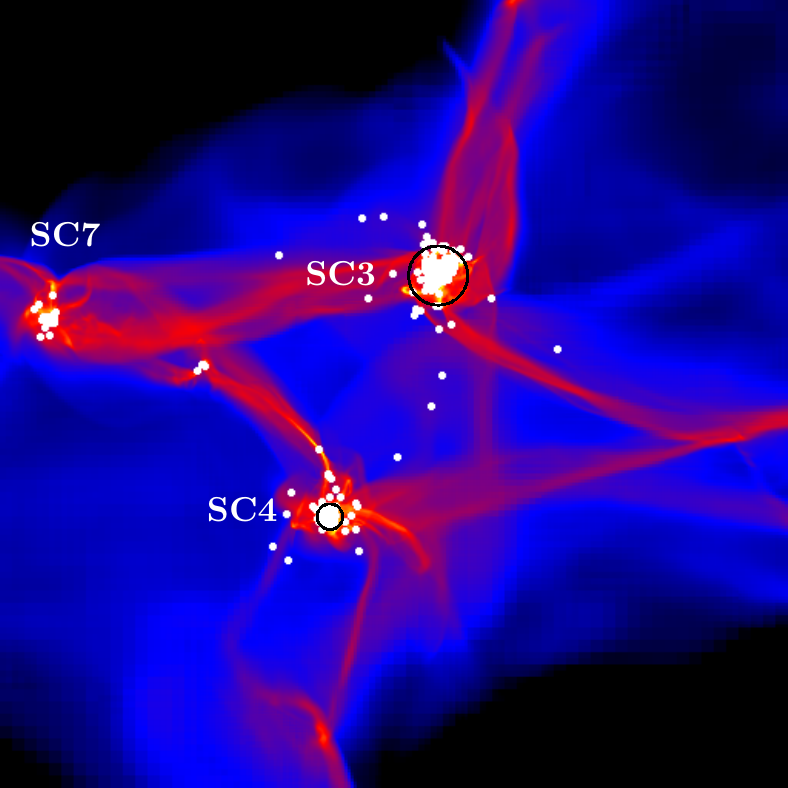}\\
    \ColorBar
  \end{minipage}
  \caption{Subclusters in the TH runs. The left picture shows TH-m-1 with the subclusters SC1 and SC2. The two largest subclusters in TH-m-2 on the right are labelled SC3 and SC4. The circles indicating the subclusters' diameter are to scale. The total size of the plot is $0.13\,\mathrm{pc}$ in both x and y direction.}
  \label{fig:TH-subclusters}
\end{figure*}

Both setups with initially uniform density distribution show distinct subclusters as illustrated in figure~\ref{fig:TH-subclusters}. We selected the four biggest subclusters for further analysis and named them SC1-SC4. The other subclusters have too few protostars for a statistical analysis. Note that subcluster SC1 is not very compact in the centre. Therefore, our reduction algorithm does not exclude the outliers, which yields the relatively large radius.

\begin{figure*}
  \begin{minipage}{180mm}
    \includegraphics[width=8cm]{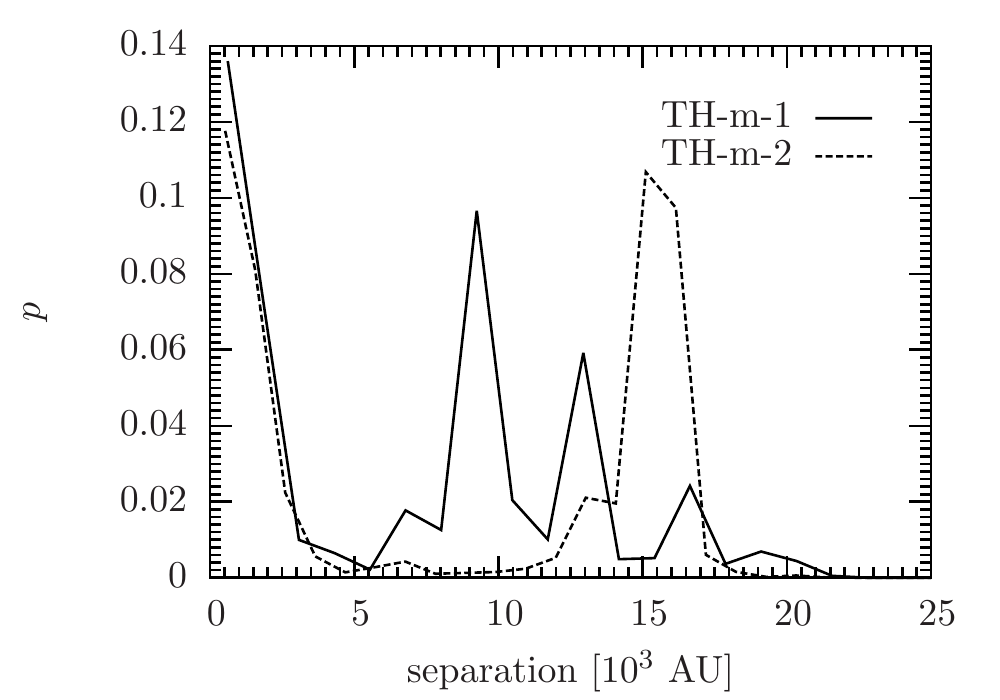}
    \includegraphics[width=8cm]{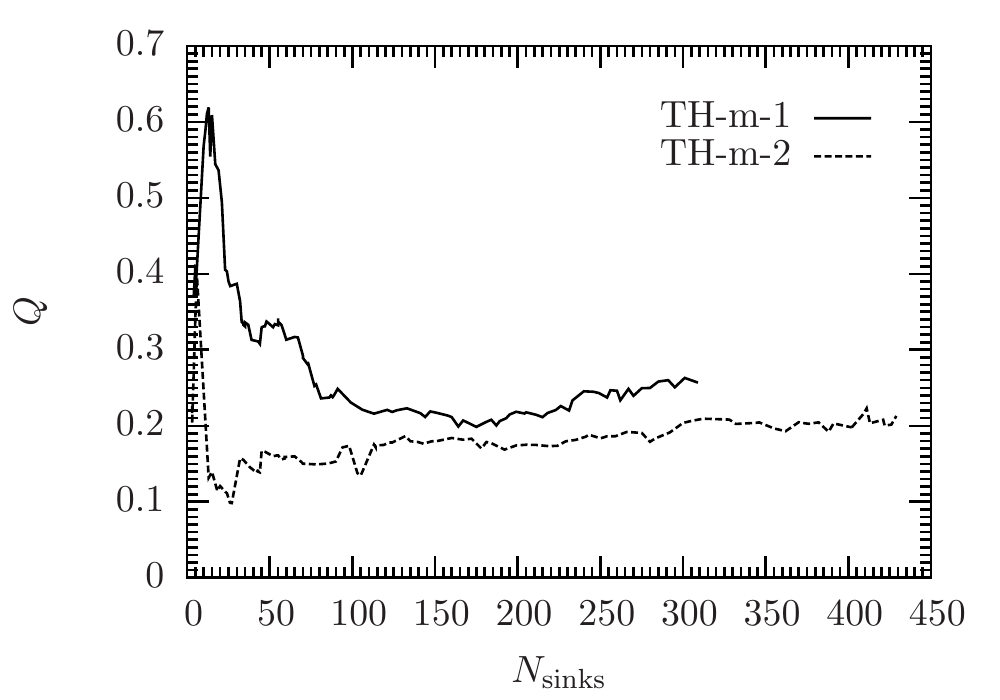}
    \caption{Global cluster values for the TH runs. Left plot: TH-m-1 shows three different peaks in the distribution function (see equation~\ref{eq:separation-distribution-function}): the one at $9{,}000~\mathrm{AU}$ corresponds to the distance of SC2 to SC1 (see figure~\ref{fig:TH-subclusters}), the peak at $13{,}000~\mathrm{AU}$ to the degenerate distance of SC2 to SC5 and SC6 and the last peak describes the distance from SC1 to SC5 and SC6, which is also degenerate within the width of the distance bin. TH-m-2 shows two main subclusters, whose distance corresponds to the peak at $15{,}000~\mathrm{AU}$. Right plot: After roughly 100 sink particles have formed, the $Q$ value (see equation~\ref{eq:Q-value}) approaches a constant value which is similar for both of the runs, indicating a high degree of substructure in both clouds.}
  \label{fig:TH-particle-cluster}
  \end{minipage}
\end{figure*}

The distribution function of the separations between the particles as well as the $Q$-value (see equation~\ref{eq:Q-value}) of the entire cloud is shown in figure~\ref{fig:TH-particle-cluster}. TH-m-1 shows three different peaks in the distribution function (see equation~\ref{eq:separation-distribution-function}): the one at $9{,}000~\mathrm{AU}$ corresponds to the distance of SC2 to SC1, the peak at $13{,}000~\mathrm{AU}$ to the degenerate distance of SC2 to SC5 and SC6, and the last peak describes the distance from the upper subcluster SC1 to SC5 and SC6, which is also degenerate within the width of the distance bin. TH-m-2 shows two main subclusters corresponding to the peak at $15{,}000~\mathrm{AU}$ in the plot. The degenerate distance between SC3 and SC7 as well as SC4 and SC7 can be seen as small peak in the distribution at $13{,}000~\mathrm{AU}$. The $Q$ value of the entire cloud shows strong variations at the beginning of stellar formation due to the different regions of the cloud where the sink particles are created. Having established the subclusters, $Q$ shows roughly constant behaviour at a value of $Q\sim0.2$ for both runs.

The key properties for the subclusters SC1--SC4 are listed in table~\ref{tab:TH-subclusters}. The protostars in SC1 have significantly larger mean separations between one another and a $Q$ value of $\sim0.7$, slightly lower than the threshold value to sub-structure of $0.8$. The other three subclusters have very similar $Q$, indicating a smooth stellar distribution.

\begin{table}
  \caption{Subcluster properties from the TH setups}
  \label{tab:TH-subclusters}
  \begin{tabular}{lcccc}
    subcluster & $N_\text{sink}$ & $\skl{s}$ [$10^3$ AU]   & $\skl{m}$ [$10^3$ AU]  & $Q$ \\
    \hline
    SC1 &  67 & 1.13 & 0.31 & 0.69\\
    SC2 & 182 & 0.49 & 0.19 & 1.36\\
    SC3 & 232 & 0.51 & 0.16 & 1.19\\
    SC4 & 100 & 0.23 & 0.10 & 1.27\\
    \hline
  \end{tabular}
    
  \medskip
  For each subcluster the number of sink particles $N_\text{sink}$, the mean separation $\skl{s}$, the mean MST length $\skl{m}$ and the $Q$ value are shown. SC1 shows signs of sub-structure indicated by a $Q$ value slightly below the critical transition value of $0.8$. SC2, SC3 and SC4 have values of $Q\gtrsim1.2$ which indicates a smooth internal structure.
\end{table}

\subsubsection{BE runs}

\begin{figure*}
  \begin{minipage}{180mm}
    \includegraphics[width=8cm]{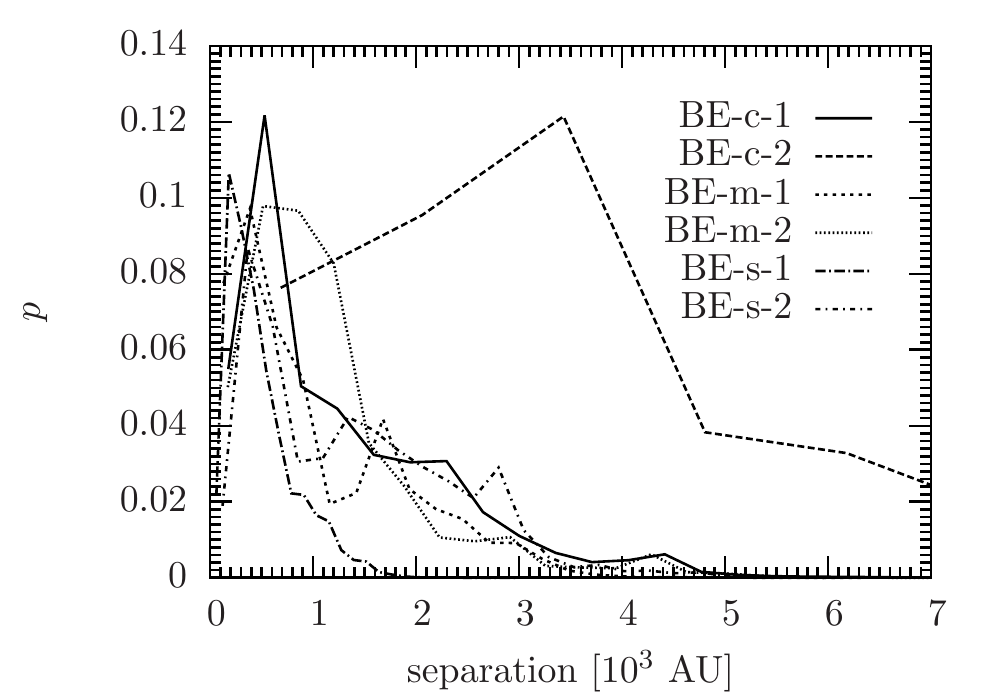}
    \includegraphics[width=8cm]{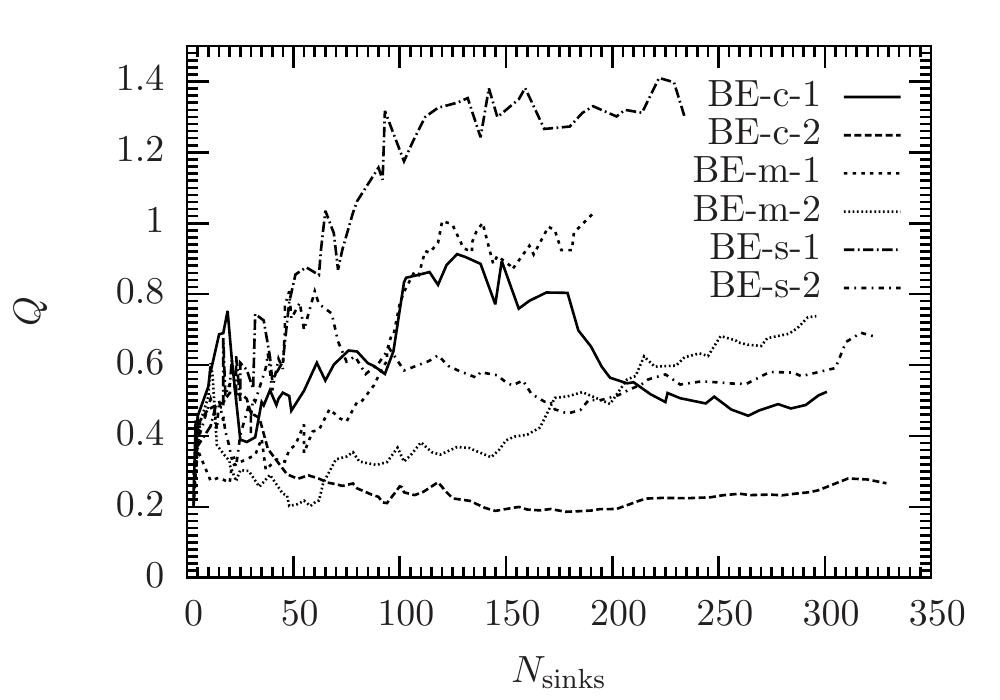}
  \caption{Cluster properties for the protostars in the BE runs. The plot of the separations (left figure) clearly shows the formation of only one main cluster for all runs, indicated by only one main peak in the distribution of protostellar separations. However, the cluster structure varies significantly (right figure). The $Q$ value differs by a factor of more than 5 for the individual runs and shows a correlation with the turbulent modes. Compressive modes show more substructure than mixed and solenoidal modes.}
  \label{fig:BE-particle-cluster}
  \end{minipage}
\end{figure*}

The effects of the much more dominant central infall during the collapse of the BE setups can be seen in the average distance between the sink particles and the $Q$-value in figure~\ref{fig:BE-particle-cluster}. The separation distribution shows only one significant maximum for all simulations. However, the peak for the BE-c-2 run is at a much larger distance. There, the sink particles form along large elongated filaments and lead to larger mean separations than in the other BE setups. Here, the strong effects of the compressive turbulent motions have a major impact. The mean separation for both runs with compressive turbulence is significantly larger than for the other runs (see $\skl{s}$ in table~\ref{tab:simulation-global-overview}). The $Q$ values and the resulting degree of substructure are very different and strongly change with time (and consequently $N_\mathrm{sink}$) depending on where the sink particles form. BE-c-2 shows strong substructure from the very beginning, BE-c-1 forms protostars at larger radii at a later stage in the simulation, leading to a decrease of $Q$ at around $N_\mathrm{sink}\sim170$.
The two runs with the lower number of sink particles (BE-s-1 and BE-m-1) have the highest values, revealing a rather smooth cluster without much substructure.

\subsubsection{PL15 runs}
\begin{figure*}
  \begin{minipage}{180mm}
    \includegraphics[width=8cm]{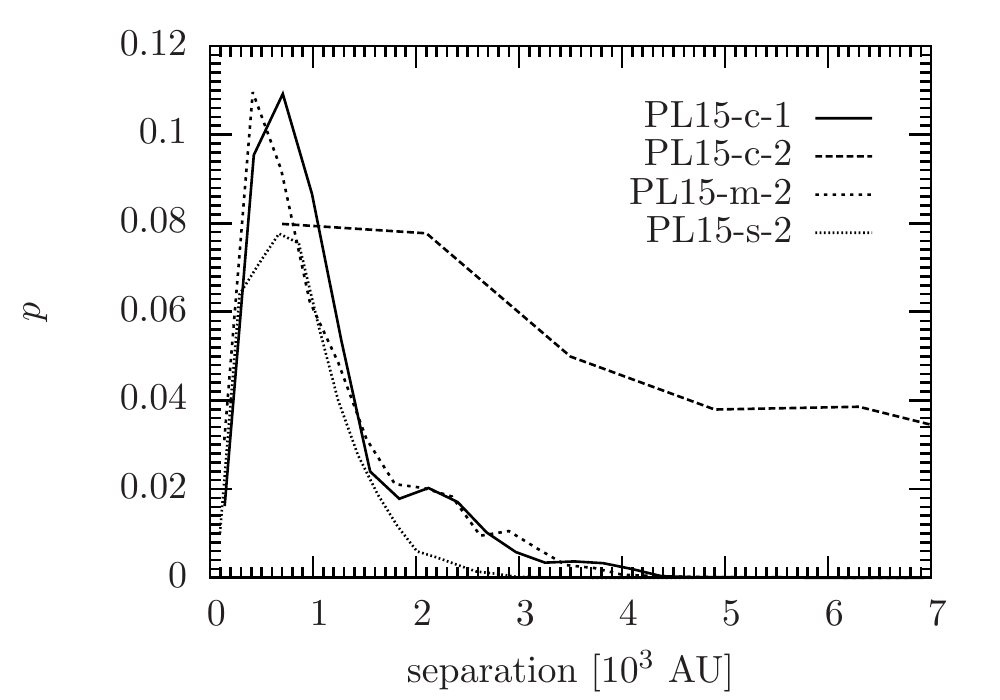}
    \includegraphics[width=8cm]{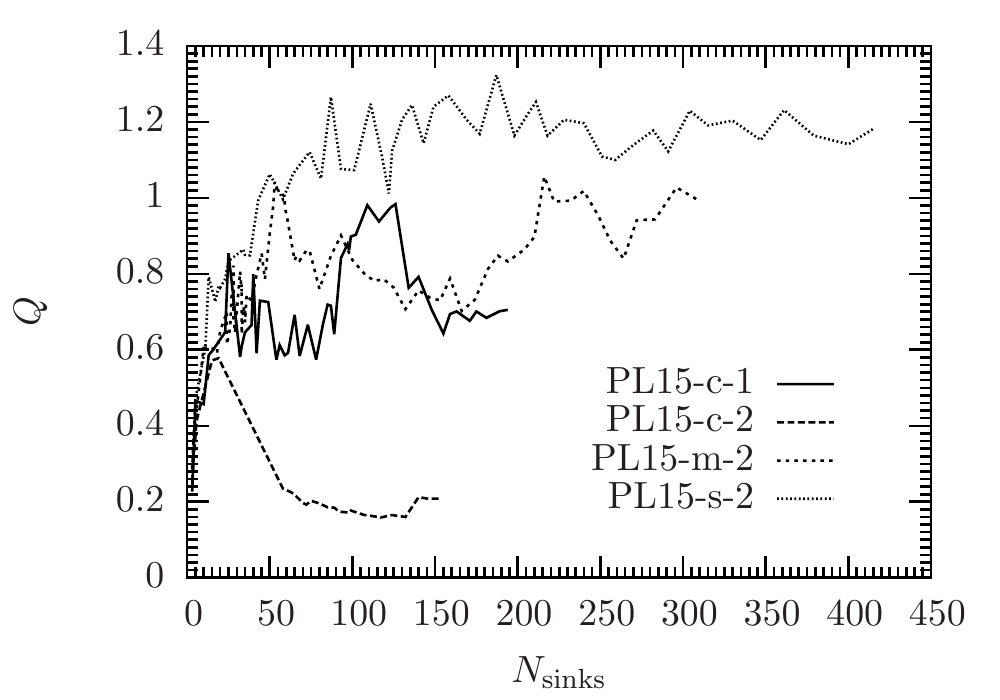}
    \caption{Cluster properties for the protostars in the PL15 runs. The plot of the separations (left figure) clearly shows the formation of only one main cluster for all runs. However, the cluster structure varies significantly (right figure).}
    \label{fig:PL15-particle-cluster}
  \end{minipage}
\end{figure*}

The even stronger mass concentration in the PL15 profiles shows a systematic influence on the mean distance between the sink particles. The mean particle separation for the PL15-c-1, PL15-m-2 and PL15-s-2 runs is roughly $15-35\%$ smaller than in the corresponding BE runs (see table~\ref{tab:simulation-global-overview}). The fact that the mean separation in PL15-c-2 is larger than in BE-c-2 is just due to the fact that the former one forms fewer sink particles; the positions of the distant sink particles at large radii are similar. Figure~\ref{fig:PL15-particle-cluster} shows the separation function and the $Q$ values. The distribution function on the left shows one main peak for all setups. The peak for PL15-c-2 is much wider, reflecting the larger central cluster. In addition, the setup forms more protostars further out than other setups. In combination with the lower total number of particles than in the BE-c-2 case, this yields the large value of $\skl{s}$ and result in the lowest $Q$ value for PL15-c-2. PL15-c-1 and PL15-m-2 are around the threshold value to substructure ($Q=0.8$), PL15-s-2 is smooth over almost all the simulated time.

\subsubsection{Comparison}
There are some general trends of the subclustering properties. The flatter the initial density profile is, the more impact has the turbulent velocity field. This causes collapsing regions to form at larger separations from each other. The observed relation $\skl{Q_\mathrm{TH}}\lesssim\skl{Q_\mathrm{BE}}\lesssim\skl{Q_\mathrm{PL15}}$ supports this intuitive picture. In a similar manner, compressive turbulent modes lead to collapsing filaments more quickly, not allowing the gas to assemble as close to the centre as in solenoidal turbulent cases. Therefore, within one density profile, the impact of turbulent modes shows $\skl{Q_\mathrm{comp}}\lesssim\skl{Q_\mathrm{mix}}\lesssim\skl{Q_\mathrm{sol}}$.

\subsection{Reduced Cluster Properties}
\label{sec:reduced-cluster}
Having analysed the total set of protostars in the entire cloud, we now focus on the central regions of the main clusters in each setup, ignoring the outliers that do not belong to the main cluster. In order to find the individual compact clusters, we iteratively exclude outlier protostars until we reach a converged cluster configuration. We first select the main region by eye. In the two TH runs we select the already mentioned subclusters (see figure~\ref{fig:TH-subclusters}), in all other setups with many sink particles we chose the central cluster. The particle reduction method works as follows. We find the centre of mass of the set of particles. Then we compute the average separation $\skl{s}$ between protostars and remove all objects that are located at radii larger than three times the mean separation from the centre of mass. We then recalculate the centre of mass and repeat the exclusion until no further particle is excluded from the set of objects. The radius of the cluster $R_\mathrm{C}$ is set to $3\skl{s}$, ensuring that all selected particles are within the cluster radius. The factor three is somewhat arbitrary, but after some tests it turned out to be a useful distance factor that does exclude all very distant particles, but no or very few particles that could be dynamically important for the cluster within the simulated time. The key values for the reduced clusters are listed in table~\ref{tab:reduced-cluster-properties}. For the following discussion we focus on the reduced clusters.

As the motions in the forming cluster are highly chaotic and the number of protostars is constantly growing, the time evolution of the reduced cluster properties fluctuates, i.e., every time step, the reduction algorithm chooses different protostars to belong to the reduced cluster. It is therefore impossible to follow single protostars within the reduced clusters. In the further analysis we thus concentrate on the clusters at the end of the simulation.

\begin{figure}
  \centering
  \includegraphics[width=8cm]{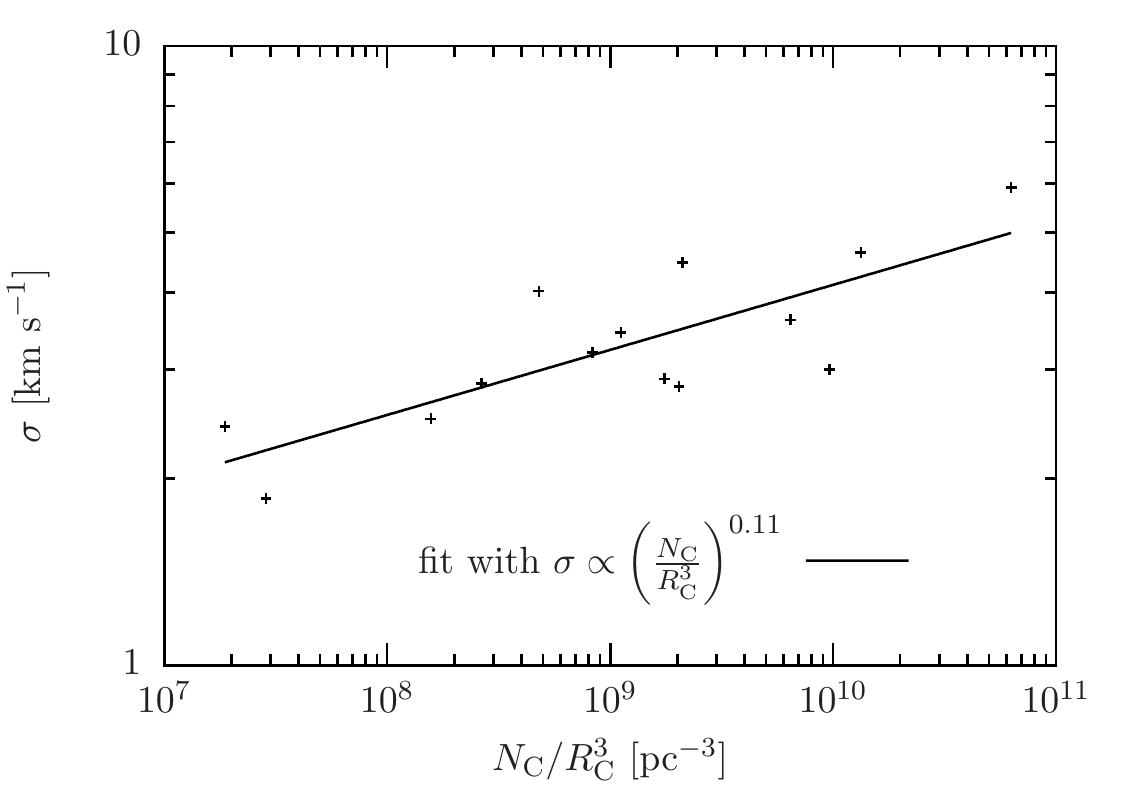}
  \caption{Velocity dispersion of the selected main (sub)clusters as a function of stellar density. The data points represent the clusters at the end of the simulation. The clusters show a weak correlation with a significant scatter.}
  \label{fig:sigma-sink-density-relation}
\end{figure}

\subsection{Mass Segregation}
We address the mass segregation problem in two ways. Firstly, we investigate the time that each protostar had for dynamical mass segregation after its formation, and secondly, we analyse the reduced cluster at the end of the simulation with the minimal spanning tree, neither taking into account the different formation times of the particles nor the change in mass during the accretion process.

\begin{figure}
  \centering
  \includegraphics[width=8cm]{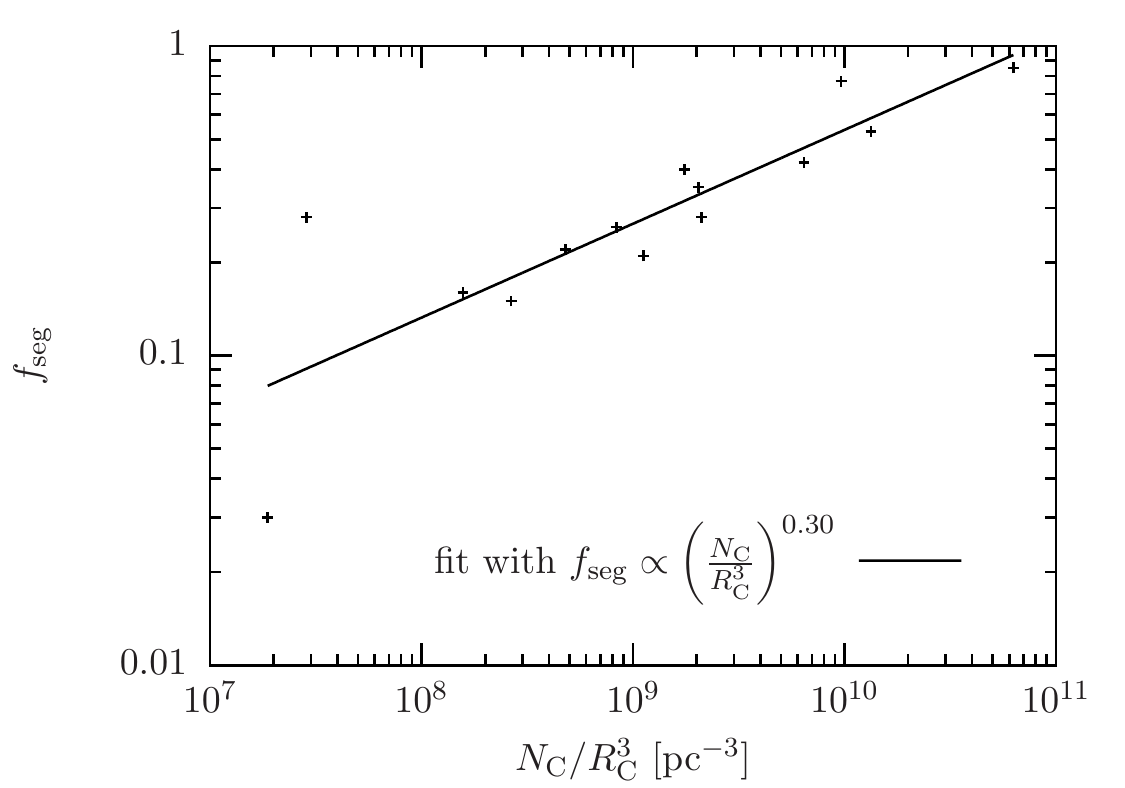}
  \caption{Possible fraction of dynamically mass-segregated stars as a function of protostellar number density of the cluster. The data points represent the reduced clusters at the end of the simulation and show a strong correlation with only little scatter.}
  \label{fig:fseg-sink-density-relation}
\end{figure}

Although the degree of mass segregation can not be calculated for a single particle but has to be seen as a global cluster property, we analyse the possibility to dynamically mass segregate via two-body relaxation for every single protostar. According to equation~(\ref{eq:mass-segregation-time}) we set the time $t_\mathrm{seg}$ to the time that the sink particle had for mass segregation, i.e., the difference between the end of the simulation and the creation time of the protostar in question. From that we infer the threshold mass $M_\mathrm{seg}$ with the given final values of $R_\mathrm{C}$ and $\sigma$. If the mass of this particular protostar is larger than the threshold mass, we count it for possible mass segregation. The quantity $N_\mathrm{seg}$ in table~\ref{tab:reduced-cluster-properties} refers to the total number of possibly mass-segregated objects; $f_\mathrm{seg}$ denotes the fraction $N_\mathrm{seg}/N_\mathrm{C}$. The strong dynamical effects during the formation of the cluster result in significantly varying values for $R_\mathrm{C}$, $N_\mathrm{C}$, and $\sigma$. However, the combined quantity in equation~(\ref{eq:mass-segregation-time}) differs much less and serves as a remarkably stable estimate. With a roughly constant formation rate of protostars, a strong correlation between the protostellar number density ($N_\mathrm{C}/R_\mathrm{C}^3$) and $f_\mathrm{seg}$ as found in the simulated clusters is not surprising (see figure~\ref{fig:fseg-sink-density-relation}). The segregation fraction $f_\mathrm{seg}$ covers a very large range ($0.03-0.85$), indicating that in some setups almost all objects had enough time to dynamically mass segregate, while in others hardly any protostar can relax in the cluster.

\begin{figure}
  \centering
  \includegraphics[width=8cm]{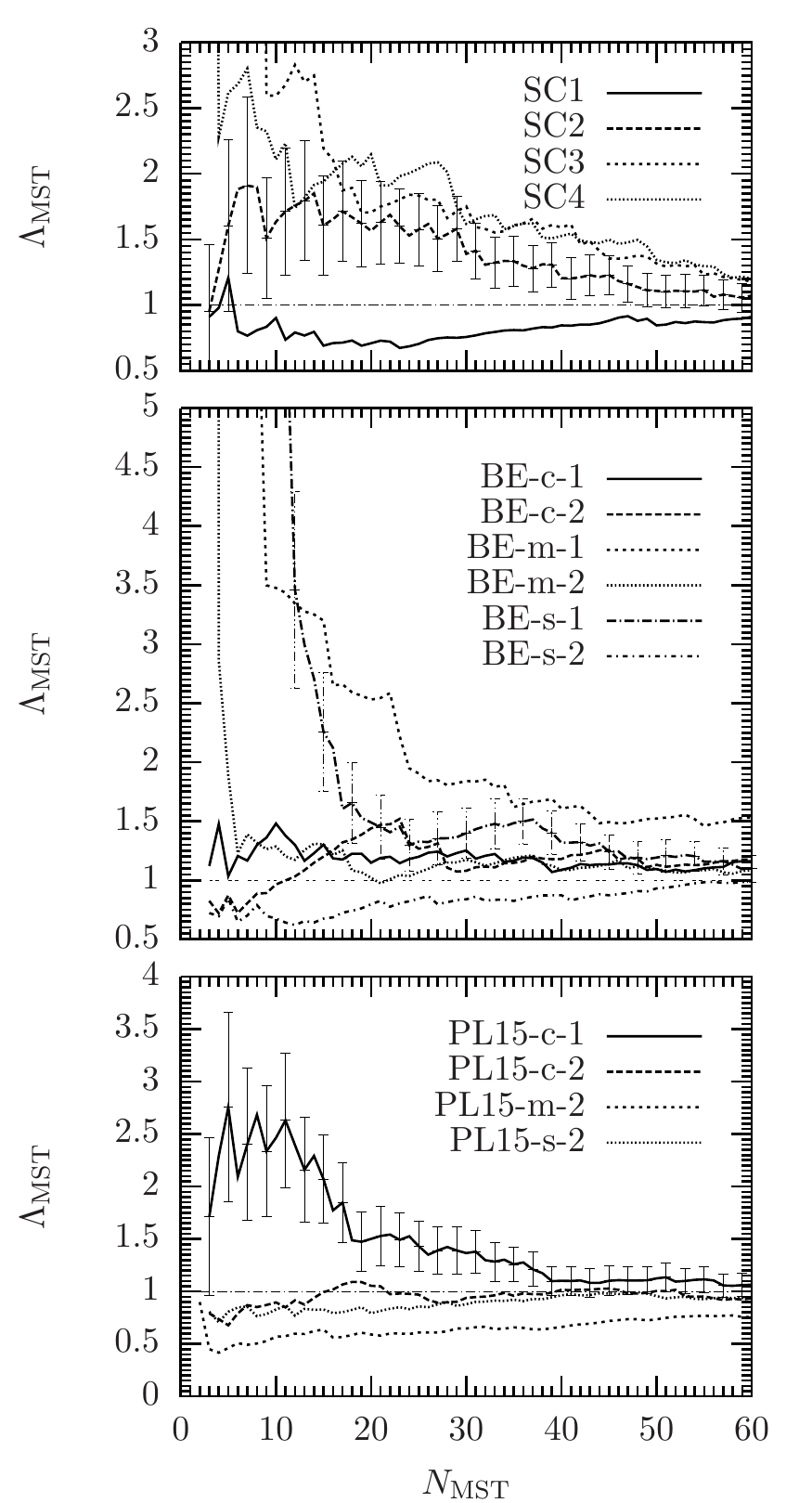}
  \caption{Mass segregation $\Lambda_\mathrm{MST}$ for all setups at the end of the simulation. For the setups where the deviation from unity is not obvious, $\Lambda_\mathrm{MST}$ is plotted with errorbars.}
  \label{fig:mass-segregation}
\end{figure}

For the second approach, we analyse the mass segregation at the end of the simulation according to equation~(\ref{eq:mass-segregation-ratio}). The values for $\Lambda_\mathrm{MST}$ as a function of $N_\mathrm{MST}$ for all clusters are plotted in figure~\ref{fig:mass-segregation}. In order to keep the plots readable, most of the curves are shown without errorbars; we included errorbars for the lowest curves that still differ from unity within a 1-$\sigma$ error in order to give some indication of the uncertainties involved. In order for mass segregation to be eminent, $\Lambda_\mathrm{MST}$ needs to be significantly above unity for mass segregation and significantly below unity for inverse mass segregation. The upper panel shows $\Lambda_\mathrm{MST}$ for the TH runs. All subclusters except for SC1 show mass segregation up to at least $N_\mathrm{MST}\sim30$, i.e., the $30$ most massive protostars form a compact subset of the cluster members around the centre of the cluster. Including more than the $30$ most massive objects to the subset enlarges the spatial extent such that the position of the chosen subset is hardly distinguishable from a random subset of the same number of cluster members. SC3 and SC4 show a significantly higher degree of mass segregation below $N_\mathrm{MST}\sim20$ and $N_\mathrm{MST}\sim12$, respectively. This corresponds to a minimum segregated sink mass of $0.074~M_\odot$ in SC3 and $0.11~M_\odot$ in SC4 and contains roughly $40\%$ and $37\%$ of the total cluster mass. Even higher values for $\Lambda_\mathrm{MST}$ can be found in the BE setups (middle panel). Here the central clusters in BE-m-1 and BE-s-1 show $\Lambda_\mathrm{MST}>1.5$ below $N_\mathrm{MST}\sim35-45$ and $N_\mathrm{MST}\sim20$, respectively. The minimum segregated mass in BE-m-1 is $M_\mathrm{seg}\approx0.1~M_\odot$, the total confined mass down to this mass is about $75\%$, in the latter case $M_\mathrm{seg}=0.17~M_\odot$, containing around $40\%$ of the cluster mass. If one includes the second bump of $\Lambda_\mathrm{MST}$ between $20<N_\mathrm{MST}<40$ in BE-s-1, the measured contained mass that is segregated is roughly $58\%$. 
Among the PL15 density profile only one cluster shows significant mass segregation, PL15-c-1. $\Lambda_\mathrm{MST}$ is greater than $1.5$ for $N_\mathrm{MST}\lesssim19$. This gives a minimum segregated mass of $M_\mathrm{seg}=0.11~M_\odot$ and corresponds to a fraction of about $72\%$ of the cluster mass.

There is a weak correlation between the actual measured mass segregation and the theoretically possible fraction of segregated protostars ($f_\mathrm{seg}$).
The actual number of segregated stars $N_\mathrm{MST,max}$ with $\Lambda_\mathrm{MST}(N\le N_\mathrm{MST,max})\gtrsim1.5$ is lower in almost all cases, but follows a consistent trend with $N_\mathrm{seg}$. Taking into account that the protostars form at different positions and need some time to dynamically relax within the cluster, the relation $N_\mathrm{MST,max}<N_\mathrm{seg}$ seems reasonable in comparison to an initially spherical cluster with a constant number of members.

The actual mass segregation can also be compared to the total time that the cluster as a whole has for mass segregation. As the number of protostars changes with time, we count the \emph{available} time starting at the point where two sink particles are formed until the end of the simulation. The ratio $t_\mathrm{avail}/t_\mathrm{relax}$ in table~\ref{tab:reduced-cluster-properties} indicates how many mass segregation times the cluster evolved, again assuming that $t_\mathrm{relax}$ at the end of the simulation is representative for the entire cluster evolution. There is again a weak correlation between this ratio and the degree of mass segregation.

\section{Discussion}
\label{sec:discussion}

In all simulations we set up cores with a very low ratio of kinetic to gravitational energy, i.e., the clouds are strongly bound. As the cores are isolated, they are disconnected from any potential dynamical impact from the surrounding environment. The initially imposed supersonic turbulent motions result in a global velocity dispersion for the gas of $\sim 0.5\,\mathrm{km\,s}^{-1}$. Given the diameter of the core, $0.2\,\mathrm{pc}$, this is close to the velocity dispersion we expect from Larson's relation \citep{Larson81,SolomonEtAl1987,Ossenkopf02,Heyer04,RomanDuvalEtAl2011}.
However, it is lower than the observed turbulent velocity component of the massive dense cores in Cygnus~X \citep{BontempsEtAl2010, CsengeriEtAl2011}. The observed cores with very similar key properties to our cores, i.e., mass, size, and temperature, show velocity dispersions from $\sim 0.5-3.5\,\mathrm{km\,s}^{-1}$, higher than the turbulent velocity dispersions in our numerical setups. Observations of massive, dense filaments show supersonic infall motions \citep{SchneiderEtAl2010}, which may easily lead to a much more dynamical formation of the cores. We do not impose an initial net rotation to the core, however, the random turbulent pattern of high and low-velocity regions in different density environments results in a net rotation of the cores with a ratio of rotational to gravitational energy ranging from $10^{-10}-10^{-3}$, in agreement with the values for the dense cores in Cygnus~X \citep{BontempsEtAl2010, CsengeriEtAl2011}. During the simulation, the velocity dispersion increases significantly due to the strong global infall and reaches values that are more consistent with the observed ones. After some $\sim10-40\,\mathrm{kyr}$, depending on the initial density profile, the cores as a whole reach or exceed a virialised energy budget $E_\mathrm{kin}/|E_\mathrm{pot}|\ge0.5$. The final energy balance is in agreement with the theoretical virial analysis in \citet{Shetty10}. They investigated the scaling relations between mass, size, and virial state of clumps of different sizes that formed self-consistently in turbulent flows. The virial state of their clumps with similar sizes and masses to our setups is consistent with our energy analysis. Also the measured line widths of our cores is consistent with the analysis in \citet{Shetty10}. The increasing values for $\sigma_\mathrm{1D}$ are dominated by the gas motions in the dense central region, which is also observed. \citet{CsengeriEtAl2011} notice small-scale turbulent motions with high velocities (a few $\mathrm{km\,s}^{-1}$) in high-resolution studies of the central region of the cores.

As soon as protostars form, the question of early substructure and mass segregation arises. These two properties of young stellar clusters can not be disentangled and analysed separately. In particular, the determination whether a cluster shows primordial or dynamical mass segregation sensitively depends on the definition of mass segregation and spatial demarcation of the region in question.

The theoretical analysis of a self-gravitating $N$-body system predicts dynamical mass segregation via two-body relaxation and dynamical friction that an object experiences while moving through a sea of other objects. For different properties of the cluster, the dynamical friction and the resulting dynamical relaxation time of the total cluster differs \citep{Chandrasekhar43,McMillanPortegiesZwart2003,SpinnatoEtAl2003,FellhauerLin2007}. The global relaxation time, defined as a statistical quantity with only global cluster properties and thus not reflecting any substructure, therefore only serves as a rough estimate. Depending on how well these global quantities fit the observed or simulated system, the relaxation time might differ significantly from the time scale of local dynamical interactions.

The question whether dynamical mass segregation can be excluded based on a time-scale argument, can therefore only be answered for a specific definition of mass segregation and for a well-defined cluster or subcluster region. Traditionally, numerical work started without initial mass segregation and investigated the purely dynamical aspect of the $N$-body system, without taking into account the dynamical changes of the individual $N$-body objects, like mass accretion in the early phase of cluster formation and the mass loss due to winds. Recently, several prescriptions of initial mass segregation have been developed (\citealt{BaumgardtEtAl2008}, \citealt{SubrEtAl2008}, \citealt{VesperiniEtAl2009}), still investigating the cluster as a whole without local substructure.

One basic problem with the analysis of mass segregation is the definition of what mass segregation actually means. \citet{Allison09} use the minimal spanning tree (MST) of the most massive stars in comparison to the MST of random stars and thus define mass segregation as the most massive stars being located closer to each other than the same number of randomly picked stars. As long as a single cluster or a conglomeration of several individual clusters does not show massive outliers, this methods works stably. In case of massive outliers, this method needs to be slightly modified \citep{MaschbergerClarke2011,OlczakEtAl2011}. In observational studies, mass segregation is mostly defined as more massive stars being located closer to the centre of the cluster (e.g. \citealt{Hillenbrand97}, \citealt{Hillenbrand98}, \citealt{Fischer98}, \citealt{deGrijs02}, \citealt{Sirianni02}, \citealt{Gouliermis04}, \citealt{HuffStahler2006}, \citealt{Stolte06}, \citealt{Sabbi08}, \citealt{GennaroEtAl2011}, \citealt{KirkMyers2011}). However, the definition of the centre of a young star forming region with a large degree of substructure is not obvious.

One possibility to study mass segregation in resolved clusters is to investigate radial variations of the initial stellar mass function. In unresolved clusters the different inferred radii in different wavelengths may indicate mass segregation. However, in both cases, mass segregation is difficult to identify given the observational difficulties \citep[e.g., ][]{AscensoEtAl2009,PortegiesZwartMcMillanGieles2010}.

Even more difficult is the answer to the question about primordial versus dynamical mass segregation. In order for \emph{global} mass segregation to be primordial in nature, it is required that stars with a given mass $m$ must be more centrally concentrated than stars with the average stellar mass $\skl{m}$ and that the cluster must be younger than the dynamical friction time scale for that given mass $m$, i.e., the more massive stars must have formed closer to the centre. This global picture is consistent with numerical simulations \citep[e.g.,][]{KlessenBurkert2000, Bonnell06}. However, this time scale argument only holds for spherical clusters in virial equilibrium. If clusters form through mergers of smaller subclusters, these subclusters might have enough time to dynamically relax and mass segregate because of the much smaller size and the higher number of stellar encounters. The degree of mass segregation in merged clusters is significantly higher than would be expected from a global time scale analysis \citep{McMillanEtAl2007, MoeckelBonnell2009}. In addition \citet{AllisonEtAl2010} show that dynamical mass segregation is very fast even without mergers of partially mass segregated substructures. Therefore, a detailed analysis of the formation of substructure and the collapse of stars within them is crucial to fully understand the mass segregation process. The analysis of our reduced clusters and subclusters with their own dynamical and orbital centre shows that there is a weak correlation between the possible degree of segregation $f_\mathrm{seg}$ and the actual mass segregation. Given the fact that we follow the evolution of our clusters for only a very short time, it seems very likely that dynamical mass segregation can provide a significant contribution to the mass segregation within the subclusters. If, in addition, the bigger stellar cluster that formed by mergers of smaller subsystems, can inherit a reasonable degree of mass segregation of the progenitors, it becomes very difficult to exclude dynamical effects on different spatial and dynamical levels to be responsible for mass segregation of a cluster.

A further complication in the mass segregation process arises from dynamical effects due to ejected stars. \citet{YuEtAl2011} showed that removing ejected stars has an effect on measuring the mass segregation of the cluster. Likewise, the initial velocity distribution influences the segregation process. In our analysis, we do not take into account the effects of escaping stars. In fact, we do not have stars that entirely escape from the cloud. Whether this is due to the low number of objects in the central region of the cluster, the relatively short evolutionary time of the simulation or due to the gravitational softening, remains an open question. Follow-up simulations that evolve the clusters for a longer time and with accurate protostellar sizes and the resulting gravitational potential are needed to clarify this effect. Concerning the initial velocity distribution, our setups show significant differences from the simulations by \citet{YuEtAl2011}. As we follow the formation of protostars in the gas cloud, the initial protostellar velocity distribution is not a free parameter, but is inherited from the gas motions of collapsing regions. In addition, the protostars in our clusters are embedded in a dense cloud whereas the simulations by \citet{YuEtAl2011} only consider the motions of the particles without background gas. Finally, their simulated times are orders of magnitude longer than in our case.

The total cluster including all protostars shows a sub-virial energy budget, indicating that the relaxation time is larger and thus the dynamical mass segregation process of the total cloud is slower than in a virialised case. However, the central regions, where the crossing times are much smaller and stellar encounters more frequent, the $N$-body system is virialised. The central region therefore does not suffer from a dynamical delay concerning the mass segregation process. In addition, the simple analysis of dynamical mass segregation does not include the effects of gas, but only the dynamical friction due to the other stellar objects in the sample. In addition, the gas also provides dynamical friction \citep{Dokuchaev1964,RudermanSpiegel1971,RephaeliSalpeter1980,Ostriker1999,LeeStahler2011}. Due to the turbulent motions, an analytic estimate is difficult to apply in our collapsing core. Nevertheless, this additional friction helps to increase the dynamical cross sections and thus makes stellar encounters more frequent, resulting in an acceleration of the dynamical mass segregation.

An interesting aspect that weakens the effect of dynamical mass segregation is presented in recent work by \citet{ConverseStahler2011}, where they argue that low-$N$ systems with an even higher number of objects than in our clusters do not relax dynamically. If this also applies to accreting stellar systems with gaseous background, a large degree of mass segregation might not be possible in the smallest subclusters but only later after some merger events. We nevertheless do not expect dynamical relaxation to become completely irrelevant because of the low number of protostars in our clusters and subclusters.

As a remark, we want to point to recent studies by \citet{KruijssenEtAl2011}. They analysed the substructure within clusters as well as the dynamical state of the stellar cluster when gas expulsion becomes important, i.e., at a slightly later stage of the evolution of the cluster. Analysing the simulations of \citet{Bonnell03,BonnellEtAl2008}, they find that the stellar system quickly reaches a globally virialised state if the gas potential is excluded and the stellar system is followed with pure $N$-body dynamics. Their results support the evolutionary picture of the formation of protostars that we see in our simulations. New protostars that form at larger radii from the centre of the cluster in gas dominated regions have sub-virial velocities. As soon as they decouple from the gas motion and move to the central gas-poor environment, they quickly virialise. The analysis of the simulations by \citet{Bonnell03,BonnellEtAl2008} with a focus on mass segregation \citep{MaschbergerClarke2011} shows global mass segregation from very early times which continues throughout the simulation. This is also in agreement with our results. Furthermore, the degree of mass segregation is only mildly influenced during subcluster merging, which suggests that early mass segregation can survive strong dynamical impacts.

Note that the number density of protostars in the central region of the clusters is high enough for protostellar collisions to become important \citep{Baumgardt11}. This could indeed lead to changes in the stellar initial mass function. A discussion of the initial stellar mass function for all setups is presented in \citet{Girichidis11a}.

Concerning the physical processes, we chose a simple setup neglecting radiative feedback, magnetic fields, jets and outflows from the young protostars, and chemical reactions. Previous studies have shown that magnetic fields tend to reduce the fragmentation (\citealt{Ziegler05}, \citealt{Banerjee06}, \citealt{HennebelleTeyssier2008}, \citealt{WangEtAl2010}, \citealt{Buerzle11}, \citealt{Hennebelle11}, \citealt{Peters11}, \citealt{SeifriedEtAl2011}) without preventing it entirely. We therefore expect our setups to form fewer stars in a magnetised environment. Similar effects would be expected if we included radiative feedback. \citet{Bate09a}, \citet{Krumholz09}, \citet{Peters10a, Peters10b, Peters10c} found reduced fragmentation in simulations, without suppressing it entirely. \citet{CommerconEtAl2011} and \citet{PetersEtAl2011} combined both, magnetic fields and radiative feedback, finding that the complex interplay between the two processes reduces fragmentation without entirely suppressing it. Concerning mechanical feedback, \citet{DaleBonnell2008} find that winds from massive stars can slow down the star formation process, but that the time-scale on which they can expel significant quantities of mass from the cluster is of the order of 10 free-fall times, which is much longer than the simulated time of our clusters. As their cores are not as dense and unstable as ours, we expect the effects of winds to be much less significant. A different fragmentation behaviour might well show differences in the formation pattern of stars, see also the different formation modes in \citet{Girichidis11b}. The substructure within the clusters is also expected to be influenced by different physical processes. In contrast, the overall energetics seems to be dominated by the global gravitational collapse of the cloud.

\section{Summary and Conclusion}
\label{sec:conclusion}

We analysed the simulations described in \citet{Girichidis11a, Girichidis11b} with the focus on the properties of the embedded young stellar clusters. We analysed the energy evolution of the gas and the nascent cluster, computed the degree of subclustering, and quantified the mass segregation in the continuously growing clusters. Our main conclusions can be summarised as follows.

In all setups, the collapsing cloud virialises within the simulated time, which corresponds to a star formation efficiency of $20\%$. Just considering the gas, all clouds have a virial or super-virial energy budget $E_\mathrm{kin} \gtrsim 0.5|E_\mathrm{pot}|$, the runs with only one protostar have significantly higher ratios of kinetic to gravitational energy. Although the total mass of all protostars is only $20\%$ of the total cloud mass, their total kinetic energy is larger than that of the gas in the cases with multiple protostars. In contrast, the three runs with only one protostar show a smaller ratio of kinetic energy of the protostar to kinetic energy of the gas, which can be explained by the vanishing momentum impact of opposite accretion flows.
Analysing the entire stellar clusters as pure $N$-body systems, we find an overall sub-virial energy balance with $E_\mathrm{kin} \sim 0.2|E_\mathrm{pot}|$, independent of the varied initial conditions. If we concentrate on the central regions of the clusters (innermost $\sim10-30\%$ of the protostars), we find virialised conditions. This difference can be explained by the formation history of the cluster. New protostars continue forming at increasing radii from the centre of the cloud due to the lack of available gas in the central region \citep{Girichidis11b}. These protostars inherit the kinetic energy from their parental gas region, which is relatively low in comparison to their gravitational contribution, i.e., new stars form at sub-virial velocities. Soon after their formation, the protostars decouple from the gas and agglomerate in the central region, where they virialise.

The degree of subclustering strongly depends on the initial density profile. Initially uniform density allows for turbulent motions to form distinct subclusters before the global collapse can confine the gas in one cluster in the central region. With a $Q$ value of $\sim0.2$, these clouds show considerable substructure with distinct conglomeration of protostars. The stronger the initial mass concentration around the centre of the cloud, the less subclustering is found. Bonnor-Ebert-like spheres show mainly one dominant central cluster with some substructure. The considered power-law density distributions form more compact protostellar clusters with less internal structure, if they form clusters at all. In three strongly condensed setups the cloud does not fragment and forms only one protostar. In general, we find that the $Q$ parameter, used to quantify subclustering, shows the following trend: $\skl{Q_\mathrm{TH}}\lesssim\skl{Q_\mathrm{BE}}\lesssim\skl{Q_\mathrm{PL15}}$, where lower $Q$ means more substructure. We also note different subclustering trends with different turbulent modes. For a given density profile, compressive modes lead to a higher degree of substructure than mixed modes, which in turn lead to more substructure than solenoidal modes, i.e., $\skl{Q_\mathrm{comp}}\lesssim\skl{Q_\mathrm{mix}}\lesssim\skl{Q_\mathrm{sol}}$.

Focusing on the central region of the clusters, where outliers are removed from the set of protostars, roughly half of the clusters show mass segregation. The degree of mass segregation varies strongly between the clusters, however, no cluster with significant inverse mass segregation is found. Except for one cluster (PL15-m-2), the mass segregation ratio does not drop below 0.5. The mass segregation is consistent with the time for dynamical mass segregation, so all the clusters had enough time for dynamical relaxation of the most massive objects in the cluster. In the simulated collapsing cores, primordial mass segregation is not necessarily required to achieve a significant mass segregation at the end of the simulation. However, due to the ongoing formation of protostars and the increase in protostellar mass due to accretion, the cluster is exposed to continuous momentum and energy impact from the surrounding gas, which may modify the actual mass segregation behaviour in comparison to the idealised process of dynamical mass segregation via two-body relaxation. A contribution that may have a significant influence is the episodic accretion of gas as well as the fact that the protostars follow the global flow pattern of the gas they form from, before they dynamically decouple from the gas. Overall, there is no clear correlation between the initial conditions and the mass segregation in our simulated clusters.

We conclude that the kinetics of young stellar clusters do not strongly depend on the initial density profile, nor on the initial structure of the turbulent modes. This is because the nascent protostars quickly decouple dynamically from the parental filament in which they were formed. The interactions as an $N$-body system dominate the cluster motions. Continuous formation of subsequent protostars with initially sub-virial velocities lead to a globally sub-virial ($E_\mathrm{kin}/|E_\mathrm{pot}| < 0.5$) state for the majority of the protostars. Taken into account the dynamics of small subclusters with dynamical times much smaller than the dynamical time of the entire cloud, the measured degree of mass segregation is fully consistent with dynamical mass segregation, there is no need for primordial mass segregation in our simulations.

\begin{appendix}
  \section{Gravitational Force Softening}
  \label{sec:smoothing}
  We used the gravitational softening for the sink particles as described in \citet{Price07a}. The potential energy can be written as
\begin{equation}
  E_\mathrm{pot} = \sum_{i\neq j}G\,m_i m_j\,\phi(r_i-r_j, h),
\end{equation}
where $h$ is the smoothing length, which is set to the accretion radius of the sink particles $h=r_\mathrm{accr}/2$, and $\phi(r,h)$ is given by ($q=r/h$)
\begin{equation}
  \phi(r,h) =
  \begin{cases}
    h^{-1}\rkl{\frac{2}{3}q^2 - \frac{3}{10}q^4 + \frac{1}{10}q^5 - \frac{7}{5}}, & 0\le q < 1\\
    h^{-1}\rkl{\frac{4}{3}q^2 - q^3 + \frac{3}{10}q^4 - \frac{1}{30}q^5 - \frac{8}{5} + \frac{1}{15q}}, & 1\le q < 2\\
    -1/r, & 2\le q.
  \end{cases}
\end{equation}
Note that the function $\phi$ is defined such that the potential energy is negative.

\section*{Acknowledgement}
P.G.~acknowledges supercomputer grants at the J\"{u}lich supercomputing centre (NIC~3433) and at the CASPUR centre (cmp09-849). P.G.~and C.F.~are grateful for financial support from the International Max Planck Research School for Astronomy and Cosmic Physics (IMPRS-A) and the Heidelberg Graduate School of Fundamental Physics (HGSFP), funded by the Excellence Initiative of the Deutsche Forschungsgemeinschaft (DFG) under grant GSC~129/1. C.F., R.B.~and R.S.K.~acknowledge financial support from the Landesstiftung Baden-W\"{u}rttemberg via their program \emph{Internationale Spitzenforschung~II} (grant P-LS-SPII/18) and from the German Bundesministerium f\"{u}r Bildung und Forschung via the ASTRONET project STAR FORMAT (grant 05A09VHA). C.F.~furthermore acknowledges funding from the European Research Council under the European Community's Seventh Framework Programme (FP7/2007-2013 Grant Agreement no. 247060) and from the Australian Research Council for a Discovery Projects Fellowship (grant no.~DP110102191). P.G. and R.B.~acknowledge funding of Emmy Noether grant BA~3706/1-1 by the DFG. R.B.~is furthermore thankful for subsidies from the FRONTIER initiative of the University of Heidelberg. R.S.K.~acknowledges subsidies from the DFG under grants no.~KL1358/10, and KL1358/11, the Sonderforschungsbereich SFB~881 \emph{The Milky Way System} as well as from a Frontier grant of Heidelberg University sponsored by the German Excellence Initiative. This work was supported in part by the U.S.~Department of Energy contract no.~DEAC-02-76SF00515. The FLASH code was developed in part by the DOE-supported Alliances Center for Astrophysical Thermonuclear Flashes (ASC) at the University of Chicago.

\end{appendix}

\clearpage
\newpage
\bibliographystyle{apj}
\bibliography{astro.bib}

\end{document}